\documentclass[conference]{IEEEtran}
\usepackage{graphicx}
\usepackage{amsmath}
\usepackage{amssymb}
\usepackage{subfigure}
\usepackage{tabu}
\usepackage{cite}
\usepackage{indentfirst}
\usepackage[utf8]{inputenc}
\usepackage[english]{babel}
\usepackage{bbm} 

\usepackage{epstopdf}
\usepackage{amsxtra}
\usepackage{amstext}
\usepackage{amsthm}
\usepackage{latexsym}
\usepackage{dsfont} 
\usepackage{xcolor}
\usepackage{units}
\usepackage{multirow}
\usepackage{lipsum}
\usepackage{bm}
\usepackage{multicol}
\usepackage{setspace}
\usepackage{tikz}
\usetikzlibrary{arrows}

\definecolor{darkgreen}{rgb}{0,0.75,0}
\usepackage[colorlinks=true]{hyperref}
\hypersetup{
	pdfauthor = {Asmaa W. Abdallah},
	pdfcreator = {Asmaa W. Abdallah},
	pdfnewwindow,
	pdffitwindow=true,
	colorlinks=true,
	linkcolor=blue,       
	citecolor=darkgreen,      
	filecolor=red,        
	urlcolor=cyan
}

\newtheorem{theorem}{Theorem}
\newtheorem{corollary}{Corollary}
\newtheorem{lemma}{Lemma}

\makeatletter
\let\NAT@parse\undefined
\makeatother

\theoremstyle{remark}

\theoremstyle{definition}

\usepackage{algorithm}
\usepackage[noend]{algpseudocode}
\algnewcommand{\LineComment}[1]{\State \(\triangleright\) #1}

\newcommand{\nth}[1]{{#1}{\text{th}}}
\newcommand{\mbf}[1]{\mathbf{#1}}

\newcommand{\norm}[1]{\left\|{#1}\right\|}

\newcommand{\expec}[1]{{\mathbb{E}\!\left[{#1}\right]}}

\renewcommand{\a}[0]{\alpha}

\renewcommand{\c}[0]{\gamma}

\newcommand{\tran}{\mathsf{T}}

\renewcommand{\Re}[1]{\mathfrak{R}\!\left\{ #1 \right\}}

\newcommand{\normm}[1]{\|{#1}\|}

\newcommand{\opt}{\mathrm{opt}}

\newcommand{\Nt}{N}
\newcommand{\Nr}{M}
\newcommand{\MbyN}{\Nr\!\times\!\Nt}
\newcommand{\NbyN}{\Nt\!\times\!\Nt}
\newcommand{\NbyM}{\Nt\!\times\!\Nr}
\newcommand{\Es}{E_{\mathrm{s}}}

\newcommand{\C}[0]{\mathcal{C}}
\newcommand{\X}[0]{\mathcal{X}}
\newcommand{\x}{\mbf{x}}
\newcommand{\xh}{\mbf{x}^{\dag}}
\newcommand{\y}{\mbf{y}}
\newcommand{\yh}{\mbf{y}^{\dag}}
\newcommand{\yt}{\tilde{\mbf{y}}}
\newcommand{\ya}{{\y}_{\mathrm{a}}}
\newcommand{\yah}{{\y}_{\mathrm{a}}^{\dag}}

\newcommand{\W}{\mbf{W}}

\newcommand{\Wp}{\mbf{W}_{\!\!\mathrm{p}}^{}}                  
\newcommand{\Wph}{\mbf{W}_{\mathrm{p}}^{\dag}}
\newcommand{\Wpi}{\mbf{W}_{\!\mathrm{p}}^{-1}}                  

\renewcommand{\H}{\mbf{H}}
\newcommand{\Hh}{\H^{\dag}}
\newcommand{\Hr}{ \H_{\mathrm{r}} }

\newcommand{\Ha}{\H_{\mathrm{a}}^{}}
\newcommand{\Hah}{\H_{\mathrm{a}}^{\dag}}

\newcommand{\I}{\mbf{I}}

\newcommand{\Q}{\mbf{Q}}
\newcommand{\Qh}{\mbf{Q}^{\dag}}
\newcommand{\Qa}[1]{\mbf{Q}_{\mathrm{a}#1}^{}}
\newcommand{\Qah}[1]{\mbf{Q}_{\mathrm{a}#1}^{\dag}}

\renewcommand{\L}{\mbf{L}}
\newcommand{\Lh}{\mbf{L}^{\dag}}

\newcommand{\Lp}{\mbf{L}_{\mathrm{p}}^{}}       
\newcommand{\Lph}{\mbf{L}_{\mathrm{p}}^{\dag}}

\newcommand{\D}{\mbf{D}}
\newcommand{\Dp}{\mbf{D}_{\mathrm{p}}^{}}

\newcommand{\Dap}{\mbf{D}_{\mathrm{ap}}^{}}

\renewcommand{\S}{\mbf{S}}

\newcommand{\Si}{\supsc{\mbf{S}}{\,-1}}

\newcommand{\Fp}{\mbf{F}_{\mathrm{p}}}
\newcommand{\Fph}{\mbf{F}_{\mathrm{p}}^{\dag}}

\newcommand{\G}{\mbf{G}}
\newcommand{\Gopt}{ \G^{\opt} }

\newcommand{\Gp}{\mbf{G}_{\mathrm{p}}}

\newcommand{\Gap}{\mbf{G}_{\mathrm{ap}}}

\newcommand{\Gapi}{\mbf{G}_{\mathrm{ap}}^{-1}}

\newcommand{\Gr}{ \mbf{G}_{\mathrm{r}} }

\newcommand{\Gropt}{ \mbf{G}_{\mathrm{r}}^{\opt} }

\newcommand{\Gi}{\mbf{G}^{-1}}

\newcommand{\La}[1]{\mbf{L}_{\mathrm{a}{#1}}^{}}       
\newcommand{\Lah}[1]{\mbf{L}_{\mathrm{a}{#1}}^{\dag}}
\newcommand{\Lai}[1]{\mbf{L}_{\mathrm{a}{#1}}^{-1}}

\newcommand{\Wmmse}{\widetilde{\mbf{W}}}            

\newcommand{\Wap}{\mbf{W}_{\!\mathrm{ap}}}                  
\newcommand{\Waph}{\mbf{W}_{\mathrm{ap}}^{\dag}}

\newcommand{\Lap}{\mbf{L}_{\mathrm{ap}}^{}}       
\newcommand{\Laph}{\mbf{L}_{\mathrm{ap}}^{\dag}}

\newcommand{\Sai}{  \subpsc{\S}{\mathrm{a}}{\,-1} }
\newcommand{\Saih}{ \subpsc{\S}{\mathrm{a}}{\,-\dag} }

\newcommand{\Fap}{\mbf{F}_{\mathrm{ap}}}
\newcommand{\Faph}{\mbf{F}_{\mathrm{ap}}^{\dag}}

\newcommand{\F}{\mbf{F}}
\newcommand{\Fr}{ \F_{\mathrm{r}} }

\newcommand{\Fropt}{ \F_{\mathrm{r}}^{\opt} }

\newcommand{\Fh}{\F^{\dag}}
\newcommand{\Fopt}{ \F^{\opt} }
\newcommand{\Fopth}{ \F^{\opt\dag} }

\newcommand{\ILB}{I_{\mathrm{LB}}}

\newcommand{\Jopt}{ \J^{\opt} }
\newcommand{\Jopth}{ \J^{\opt\dag} }

\newcommand{\J}{\mbf{J}}
\newcommand{\Jh}{\J^{\dag}}

\newcommand{\U}{\mbf{U}}

\newcommand{\Uh}{\mbf{U}^{\dag}}
\newcommand{\Uopt}{ \U^{\opt} }

\newcommand{\V}{\mbf{V}}
\newcommand{\Vh}{\mbf{V}^{\dag}}

\newcommand{\E}{\mbf{E}}
\newcommand{\skk}{\subpsc{s}{kk}{}}

\newcommand{\yp}{\mbf{y}_{\mathrm{p}}}

\newcommand{\Wmmseh}{{\widetilde{\mbf{W}}}^{\dag}}    

\newcommand{\mup}[1]{\mu_{\mathrm{p}}\!\left(#1\right)}
\newcommand{\muap}[1]{\mu_{\mathrm{ap}}\!\left(#1\right)}
\newcommand{\mum}[1]{\mu_{\mathrm{m}}\!\left(#1\right)}

\newcommand{\Tr}[1]{\mathrm{Tr}\!\left( #1 \right)}
\newcommand{\diag}[1]{\mbox{\small$\mathrm{diag}$}\!\left( #1 \right)}

\newcommand\supsc[2]{#1^{\!\raisebox{+.3ex}{\scalebox{.8}{$\scriptstyle #2$}}}}
\newcommand\subsc[2]{#1_{\raisebox{+.1ex}{\scalebox{.8}{$\scriptstyle  #2$}}}}
\newcommand\subpsc[3]{#1_{\raisebox{+.1ex}{\scalebox{.8}{$\scriptstyle  #2$}}}^{\!\raisebox{+.3ex}{\scalebox{.8}{$\scriptstyle #3$}}}}

\newcommand{\ML}{\mbox{\tiny$\mathrm{ML}$}}
\newcommand{\WLD}{\mbox{\tiny$\mathrm{WLD}$}}
\newcommand{\AWLD}{\mbox{\tiny$\mathrm{AWLD}$}}
\newcommand{\LB}{\mbox{\tiny$\mathrm{LB}$}}
\newcommand{\ILBWLD}{I_{\LB}^{\WLD}}
\newcommand{\ILBAWLD}{I_{\LB}^{\AWLD}}

\newcommand{\Ex}[2]{\mathsf{E}_{#1}\!\left[ #2 \right]}

\DeclareMathOperator*{\argmax}{\mathrm{argmax}}

\begin{document}
	
\title{Optimal Augmented-Channel Puncturing for Low-Complexity Soft-Output MIMO Detectors
}
\vspace{-0.05in}

\author{\IEEEauthorblockN{Mohammad M. Mansour}
\IEEEauthorblockA{\textit{Dept. of Electrical and Computer Engineering} \\
\textit{American University of Beirut}\\
Beirut, Lebanon \\
mmansour@ieee.org}
}
\vspace{-0.05in}
\maketitle

\begin{abstract}
We propose a computationally-efficient soft-output detector for multiple-input multiple-output channels based on augmented channel puncturing in order to reduce tree processing complexity. The proposed detector, dubbed augmented WL detector (AWLD), employs a punctured channel with a special structure derived by triangulizing the original channel in augmented form, followed by Gaussian elimination. We prove that these punctured channels are optimal in maximizing the lower-bound on the achievable information rate (AIR) based on a newly proposed mismatched detection model. We show that the AWLD decomposes into a minimum mean-square error (MMSE) prefilter and channel-gain compensation stages, followed by a regular unaugmented WL detector (WLD). It attains the same performance as the existing AIR partial marginalization (AIR-PM) detector, but with much simpler processing.
\end{abstract}
	

%
\vspace{-0.05in}
\section{Introduction}\label{s:intro}
Multiple-input multiple-output (MIMO) communications using a large number of transmit and receive antennas has become mainstream technology in most modern wireless standards, primarily in 5G, in order to support the aggressive targets set on spectral efficiencies. However, achieving the ideal performance promised by this technology requires the use of MIMO detectors whose complexity grows exponentially in the number of transmit antennas $\Nt$ and polynomially in the size $Q$ of the signal constellation $\X$. To support low-latency communications while providing high throughput, computationally efficient designs of MIMO detectors that do not incur substantial performance loss are needed, especially for large MIMO dimensions and dense constellations.

The topic of MIMO detection is a classical area of research, and the literature is very rich with schemes that provide various performance-complexity tradeoffs in the design space (e.g., see overviews in~\cite{2018_Xu_two_decades, 2009_larsson_MIMO_detection_methods}). The benchmark for performance in the sense of generating good soft decisions on the transmitted information bits
remains the maximum likelihood (ML) detection scheme, which provides optimal performance at exponential complexity. Alternatively, the benchmarks for low-complexity are the zero-forcing (ZF) and minimum mean-square error (MMSE) schemes, which decouple the transmit layers through a linear filtering stage to generate log-likelihood ratios (LLRs) for each bit symbol either in parallel or sequentially through decision feedback. Although linear processing incurs only a marginal loss in mutual information between the transmitter and receiver, and offers fairly good performance in fast fading channels, it severely limits the diversity order of a MIMO system in slowly fading channels~\cite{2008_larsson_fixed_complexity}.

Tree-search based detectors such as sphere decoding~\cite{2003_damen_on_ML_detection}, list decoding~\cite{2003_hochwald_achieving_nearcapacity}, and other variants map the detection problem into a search problem for the closest signal vector. They find the closest $\x\!\in\!\X$ to the received vector $\y$ by forming a search-tree and recursively enumerating all symbols in $\X$ across all layers in $\x$ from the parent down to the leafs. Such schemes suffer from non-deterministic complexity (see scheduling solutions in~\cite{2014_sphereP2_mansour}). To simplify the search process, fixed-complexity schemes such as~\cite{barbero2008fixing,2006_Wenk_ISCAS} limit the search steps to a set of survivor paths. While these schemes are efficient in finding the ML path, they do not necessarily find all the best competing paths that are needed to generate soft decisions.

An alternative concept is partial marginalization (PM)~\cite{2008_larsson_fixed_complexity,2011_persson_partial_marg}, which exhaustively enumerates only over a small subset $\nu$ of carefully chosen parent layers, and marginalizes over the other $\Nt\!-\!\nu$ child layers using ZF with decision-feedback estimates. While the bit LLRs for parent symbols are easy to compute, computing bit LLRs for child symbols is complicated by two facts: 1) each child \emph{bit} requires a separate QR decomposition (QRD), totalling $Q(\Nt\!-\!\nu)$, and 2) the LLRs are prone to error propagation for large $\Nt\!-\!\nu$ due to decision feedback. In~\cite{2006_siti_novel_LORD}, the closely related layered orthogonal lattice detector (LORD) scheme mitigates the first drawback by operating with $\nu\!=\!1$ and computing bit LLRs for the parent symbol only; $\Nt$ independent QRDs and tree searches are performed to compute the bit LLRs for all symbols by choosing a different symbol as parent each round.

To overcome the second drawback, the WL detection (WLD) scheme~\cite{2014_mansour_SPL_WLD} first applies a (non-unitary) filtering matrix $\W$ to decompose the channel into a sparse lower-triangular matrix $\L$ (and hence the acronym WLD). It then enumerates across one parent layer and detects symbols in all other child layers in parallel via least-squares (LS) estimates with no decision feedback. The channel matrix is ``punctured'' to have a special structure in order to break the connections among child nodes, while retaining connections only to the parent. Essentially, all child nodes become leaf nodes, and hence LS estimates are optimal. An immediate consequence is that the LS estimates of the counter hypotheses of the $Q$ bits in each leaf symbol can be easily derived from the LS estimate itself~\cite{2014_sphereP1_mansour}. A closely related concept is the achievable information rate (AIR)-PM detector~\cite{2012_rusek_optimal_channel_short,2017_hu_softoutput_AIR}, which derives a ``shortened'' channel similar to the WLD's punctured structure using information-theoretic optimizations.

In this paper, we show that the concepts of channel puncturing of~\cite{2014_mansour_SPL_WLD} and AIR-PM-based channel shortening of~\cite{2017_hu_softoutput_AIR} are related. After introducing the system model in Sec.~\ref{s:system_model}, we first present a matrix characterization of the WLD detector of order $\nu$ in terms of Gaussian elimination matrices. We then derive a lower bound on the achievable rate of the WLD detector, as well as a bound on the quality of its hard decision estimate, and show that these bounds approach capacity and the ML hard-decision as $\nu$ increases (Sec.~\ref{sec:WLD}). We also propose a new \emph{augmented} WLD (AWLD) MIMO detection scheme in Sec.~\ref{sec:augmented_WLD} in which an augmented channel rather than the original channel is punctured. We derive a lower bound on the AIR of the AWLD detector and characterize its gap to capacity. In Sec.~\ref{s:modified_detection_model}, we propose an alternate mismatched detection model compared to~\cite{2012_rusek_optimal_channel_short}
and use it derive optimal punctured channel matrices that maximize the AIR.
We prove that the AWLD detector is optimal under this model, and is in fact equivalent to the AIR-PM detector of~\cite{2017_hu_softoutput_AIR}. The AWLD detector decomposes into an MMSE prefilter and channel-gain compensation stages, followed by an unaugmented WLD. Hence, AIR-optimal channel puncturing can be achieved using simple QR decomposition followed by Gaussian elimination.

%
\section{System Model}\label{s:system_model}
Consider a MIMO system with $\Nt$ transmit antennas and $\Nr\!\geq\!\Nt$ receive antennas. Let $\H\!\in\!\mathcal{C}^{\MbyN}$ represent the MIMO communication channel, which is assumed to be perfectly known at the receiver. The transmit signal $\x\!=\![x_n^{}]\!\in\!\X^{\Nt\times 1}$ is composed of $\Nt$ symbols $x_n^{}$ drawn from constellation $\X\!\subset\!\mathcal{C}$ with average energy $\expec{x_n^{}x_n^{\dag}}\!=\!\Es$, where each symbol $x_n^{}\!=\!(x_n^{q})$ is mapped from $Q$ bits $x_n^{q}\!\in\!\{\pm 1\}$. The receive signal $\y\!\in\!\mathcal{C}^{\Nt\times 1}$ is modeled using the input-output relation\vspace{-0.075in}
\begin{align}\label{eq:system_model}
    \y = \H\x + \mbf{z},\\[-1.75em]\notag
\end{align}
where the noise term $\mbf{z}\!\sim\! \mathcal{CN}(\mbf{0}_{\Nr\times 1}^{},N_0^{}\I_{\Nr}^{})$. The conditional probability $ p(\y|\x)$ and metric $\mu(\y|\x)$ according to~\eqref{eq:system_model} are\\[-1.5em]
\begin{align}
    p(\y|\x)
    &=
    \tfrac{1}{(\pi N_0)^N}\! \exp{\left( \mu(\y|\x) \right)}, \label{eq:true_detection_prob}
    \\
    \mu(\y|\x)
    &=
    -\tfrac{1}{N_0}\norm{\y\!-\!\H\x}^2 \label{eq:true_metric}
    \\
    &=
    -\tfrac{1}{N_0}(\yh\y \!-\! 2\Re{\yh\H\x} \!+\! \xh\Hh\H\x) \label{eq:true_metric_expanded}
    \\
    &\propto
    2\Re{\yh\H\x} \!-\! \xh\Hh\H\x. \label{eq:true_metric_expanded_sufficient}
\end{align}\\[-1.5em]
Using the observation $\y$ and assuming no prior information on $\x$, the ML detector generates the LLR of the $\nth{q}$ bit $x_n^q$ of the $\nth{n}$ symbol $x_n^{}$ in $\x$ as\\[-1.5em]
\begin{align}
    L(x_n^q|\y)
    &=
    \ln
    \frac{\sum_{\x:x_n^q=+1}\exp\left(\mu(\y|\x)\right)}
         {\sum_{\x:x_n^q=-1}\exp\left(\mu(\y|\x)\right)}.
\label{eq:LLR_ML_def}
\end{align}
To avoid computing exponentials, the approximation~\cite{2005_moon_error_correcting_codes} $\ln(e^x\!+\!e^y)\!\approx\!\max(a,b)$ can be applied to approximate as~\eqref{eq:LLR_ML_def}\vspace{-0.05in}
\begin{align}
    L(x_n^q|\y)
    &\approx
    \max_{\x:x_n^q=+1}\mu(\y|\x) -
    \max_{\x:x_n^q=-1}\mu(\y|\x).
\label{eq:LLR_MLM_def}\\[-2em]\notag
\end{align}
In the absence of any structure on $\H$, computing the sums in~\eqref{eq:LLR_ML_def} or max terms in~\eqref{eq:LLR_MLM_def} have exponential complexities.

%
\vspace{-0.05in}
\section{WLD MIMO Detector}\label{sec:WLD}
Let $\H\!=\!\Q\L$ denote the (thin) QL decomposition~\cite{2013_golub_matrix} of $\H$, where $\Q\!\in\!\mathcal{C}^{\MbyN}$ has orthonormal columns, $\L\!\in\!\mathcal{C}^{\NbyN}$ is lower-triangular with real positive diagonal elements. In~\cite{2014_mansour_SPL_WLD,2014_mansour_eurasip_WLD}, a technique to puncture $\L\!=\![l_{ij}^{}]$ into $\Lp\!\in\!\mathcal{C}^{\NbyN}$ by nulling all entries below the main diagonal and to the right of the first column ($l_{ij}^{}\!\leftarrow\!0$ for $1\!<\!j\!<\!i$ and $1\!<\!i\!<\!\Nt$) using Gaussian elimination is presented. Here, we give an alternate characterization using matrices, and generalize it to other puncturing patterns. Assume $\L$ is partitioned as\vspace{-0.075in}
\begin{align}
    \mbf{L}_{\Nt\times\Nt}^{} =
    \begin{bmatrix}
        p & \mbf{0}_{1\times(\Nt-1)}^{}  \\
        \mbf{r}_{(\Nt-1)\times 1}^{} & \mbf{S}_{(\Nt-1)\times(\Nt-1)}^{}
    \end{bmatrix}.\label{eq:L_partitioned}
\end{align}\\[-1.0em]
where $p$ is a real scalar. For non-singular $\L$, $\Lp$ is given by\vspace{-0.05in}
\begin{align}
    \Lp &\triangleq \Wp\L
        =
        \Dp \diag{\L}
        \begin{bmatrix}
            \tfrac{1}{p} & \mbf{0}  \\
            \mbf{0} & \mbf{S}^{-1}
        \end{bmatrix}
        \begin{bmatrix}
            p & \mbf{0} \\
            \mbf{r} & \mbf{S}
        \end{bmatrix}\label{eq:LP_def}
\end{align}\\[-1.0em]
The diagonal matrix $\Dp\!\in\!\mathcal{R}^{\NbyN}$ is chosen such that the puncturing matrix $\Wp\!\in\!\mathcal{C}^{\NbyN}$ satisfies $\diag{\Wp\Wph}\!=\!\I_{\Nt}$:\vspace{-0.05in}
\begin{align}
        \Dp
        &=
        \begin{bmatrix}
            1 & \mbf{0}  \\
            \mbf{0} & \supsc{\diag{\mbf{S}}}{-1}\mbf{\Sigma}
        \end{bmatrix}\label{eq:Dp_formula}
        \\
        \Wp
        &=
        \begin{bmatrix}
            1 & \mbf{0}  \\
            \mbf{0} & \mbf{\Sigma}\mbf{S}^{-1}
        \end{bmatrix}\label{eq:Wp_formula}
        \\
        \mbf{\Sigma}
        &=
        \supsc{\diag{\mbf{S}^{-1}\mbf{S}^{-\dag}}}{-1/2}. \label{eq:Sigma}
\end{align}\\[-2.5em]

The above definition of $\Lp$ can be generalized to any lower-triangular puncturing pattern of order $1\!\leq \nu\!\leq\Nt\!-\!1$ as follows:\vspace{-0.05in}
\begin{align}
    \mbf{L}_{\Nt\times\Nt}^{(\nu)}
    &=
    \begin{bmatrix}
        \mbf{P}_{\nu\times\nu} & \mbf{0}_{\nu\times(\Nt-\nu)}^{}  \\
        \mbf{R}_{(\Nt-\nu)\times \nu}^{} & \mbf{S}_{(\Nt-\nu)\times(\Nt-\nu)}^{}
    \end{bmatrix}\label{eq:L_nu_partitioned}
    \\
    \mbf{L}_{\mathrm{p}}^{(\nu)}
        &\triangleq \W_{\mathrm{p}}^{(\nu)} \L^{(\nu)}
        \!=\!
        \D_{\mathrm{p}}^{(\nu)} \diag{\L}
        \begin{bmatrix}
            \I & \mbf{0}  \\
            \mbf{0} & \mbf{S}^{-1}
        \end{bmatrix}\!\!
        \begin{bmatrix}
            \mbf{P} & \mbf{0} \\
            \mbf{R} & \mbf{S}
        \end{bmatrix}\label{eq:LP_nu_def1}
\end{align}\\[-1.5em]
where $\D_{\mathrm{p}}^{(\nu)}$ and $\W_{\mathrm{p}}^{(\nu)}$ are given by \vspace{-0.05in}
\begin{align}
        \D_{\mathrm{p}}^{(\nu)}
        &\!=\!
        \supsc{\diag{\L^{(\nu)}}}{-1}\!
        \begin{bmatrix}
            \I & \mbf{0}   \\
            \mbf{0} & \mbf{\Sigma}
        \end{bmatrix}\label{eq:Dp_nu_formula}
        \\
        \W_{\mathrm{p}}^{(\nu)}
        &\!=\!
        \begin{bmatrix}
            \I & \mbf{0}  \\
            \mbf{0} & \mbf{\Sigma}\Si
        \end{bmatrix}\label{eq:Wp_nu_formula}
\end{align}\\[-1.0em]
Note that $\W_{\mathrm{p}}^{(\nu)}$ is a non-singular lower triangular matrix with $\nu$ ones on the main diagonal. Also, since $\mbf{\Sigma}$ normalizes the diagonal elements of $\Si$, then the remaining $\Nt\!-\!\nu$ eigenvalues of $\W_{\mathrm{p}}^{(\nu)}$ are positive and less than or equal to 1. Therefore, it follows that $\sigma_{\max} \!\geq\! \lambda_{\max} \!\geq\! 1$ and $0 \! < \! \sigma_{\min} \!\leq\! \lambda_{\min} \!\leq\! 1$, where $\sigma_{\max}$ ($\sigma_{\min}$) and $\lambda_{\max}$ ($\lambda_{\min}$) are the maximum (minimum) singular values and eigenvalues of $\W_{\mathrm{p}}^{(\nu)}$, respectively. For simplicity of notation, we drop the superscript $(\nu)$, with the understanding that the puncturing order is $\nu$.
\vspace{-0.075in}

%
\hspace{0.02in}\subsection{WLD MIMO Detection Model}\label{s:WLD_detection_model}
By applying the filtering matrix $\Wp$, the metric in~\eqref{eq:true_metric} computed by the WLD detector takes the form
\begin{align}
    -\tfrac{1}{N_0}\!\norm{\Qh\y\!-\!\L\x}^2
    \xrightarrow{~\Wp~}
    -\tfrac{1}{N_0}\!\norm{\Wp(\Qh\y\!-\!\L\x)}^2. \label{eq:wld_metric_tmp}
\end{align}
Next, expanding~\eqref{eq:wld_metric_tmp} and dropping the irrelevant term $-\tfrac{1}{N_0}\norm{\Wp\Qh\y}^2$,~\eqref{eq:wld_metric_tmp} can be rewritten as
\begin{align}
    \mup{\y|\x}
    &\!=\!
    2\Re{\yh\Fp\x} \! - \! \xh\Gp\x, \label{eq:wld_metric}
\end{align}
where $\Fp \!\triangleq\! \tfrac{1}{N_0}\Q\Wph\Lp$, and $
\Gp\!\triangleq\! \tfrac{1}{N_0}\Lph\Lp = \Hh\Fp$. The corresponding detection model becomes
\begin{align}
    p_{\mathrm{p}}(\y|\x)
    &=
    \exp{(2\Re{\yh\Fp\x} \! - \! \xh\Gp\x)}, \label{eq:WLD_detection_prob}
\end{align}
instead of the true conditional probability in~\eqref{eq:true_detection_prob}. Based on~\eqref{eq:WLD_detection_prob}, the AIR of the WLD detector is lower-bounded by~\cite{2006_arnold_simulation_based}
\begin{align}
    \ILBWLD
        &= \Ex{\mbf{Y},\mbf{X}}{\log( p_{\mathrm{p}}(\y|\x) )}
            -\Ex{\mbf{Y}}{\log( p_{\mathrm{p}}(\y) )},\label{eq:ILB_WLD}
\end{align}
where the expectations are taken over the true channel statistics, and $p_{\mathrm{p}}(\y)
    \triangleq
    \int p_{\mathrm{p}}(\y|\x) p(\x)\,\mathrm{d}\x$, with $p(\x)$ being the prior distribution of $\x$.

%
\begin{theorem} Assuming $\x\!\sim\! \mathcal{CN}\!(\mbf{0},\Es\I_{\Nt})$, and let $\beta\!=\! \tfrac{E_s}{N_0}$ be the signal-to-noise ratio (SNR), then the AIR of the WLD detector is lower-bounded by\vspace{-0.05in}
\small
\begin{align*}
    \ILBWLD
        \!=\!  \log \det \!\left( \I \!+\! \beta\Lph\Lp \right) \!-\!
            \Tr{(\I \!-\! \Wp\Wph)(\I \!+\! \beta\Lp\Lph)^{-1} }\\[-3.0em]
\end{align*}\normalsize
\label{thm:ILB_wld}
\end{theorem}
\begin{IEEEproof} We compute the expectations in~\eqref{eq:ILB_WLD} as\vspace{-0.05in}
\begin{align*}
    \Ex{\mbf{Y},\mbf{X}}{\log( p_{\mathrm{p}}(\y|\x) )}
    &=
    \Es\Tr{\Gp}
    \\
    -\Ex{\mbf{Y}}{\log( p_{\mathrm{p}}(\y) )}
    &=
    \Nt\log\Es
    +
    \log\det( \Gp \!+\! \tfrac{1}{\Es}\I)
    \\
    -
    &\Tr{ \Fph\left[ \Es\H \Hh\!+\! N_0\I  \right]\! \Fp
            [ \Gp \!+\! \tfrac{1}{\Es}\I]^{-1}  }\\[-2em]
\end{align*}
following~\cite{2012_rusek_optimal_channel_short}. Substituting for $\Fp\!=\!\tfrac{1}{N_0}\Q\Wph\Lp$, and $\Gp\!=\!\tfrac{1}{N_0}\Lph\Lp$, and applying the matrix inversion lemma~\cite{2011_zhang_matrix_theory} followed by standard simplifications, the result follows.
\end{IEEEproof}

Note that for $\nu\!=\!\Nt\!-\!1$, we have $\Wp\!=\!\I$ and $\Lp\!=\!\L$, and then $ \ILBWLD
\!=\!  \log \det \!\left( \I \!+\! \beta\Lh\L \right)$, which is the capacity of the channel. In fact, as $\nu$ increases from 1, the metrics computed by the WLD detector approach the hard-decision ML metrics as shown by the following lemma:

%
\begin{lemma}\label{lem:WLD_distance_bound} Let $\x_{\ML}\!=\!\arg\min_{\x\in\X^{\Nt}}\!\!\norm{\yt\!-\!\L\x}$ and $\x_{\WLD}\!=\!\arg\min_{\x\in\X^{\Nt}}\!\!\norm{\Wp(\yt\!-\!\L\x)}$ where $\H\!=\!\Q\L,\yt\!=\!\Qh\y$, then
\begin{align}
   \norm{\yt\!-\!\L\x_{\ML}}
   \leq
   \norm{\yt\!-\!\L\x_{\WLD}}
   &\leq
   \kappa(\Wp)\!\norm{\yt\!-\!\L\x_{\ML}}\label{eq:WLD_distance_bound}
   \\
   \norm{\Wp(\yt\!-\!\L\x_{\WLD})}
   &\leq
   \sigma_{\max}{(\Wp)}\norm{\yt\!-\!\L\x_{\ML}} \label{eq:WLD_distance_bound2}
\end{align}
where $\kappa(\Wp)\!=\!\sigma_{\max}(\Wp)/\sigma_{\min}(\Wp)$ is the condition number of $\Wp$, and $\sigma_{\max}{(\Wp)},\sigma_{\min}{(\Wp)}$ are the largest and smallest singular values of $\Wp$, respectively.
\end{lemma}
\begin{IEEEproof} The first inequality in~\eqref{eq:WLD_distance_bound} follows from the definition of the ML solution. For the second, we have
\begin{align}
   \norm{\yt\!-\!\L\x_{\WLD}}
   &=
   \norm{\Wpi\Wp(\yt\!-\!\L\x_{\WLD})}\notag
   \\
   &\leq
   \sigma_{\max}(\Wpi)\norm{\Wp(\yt\!-\!\L\x_{\WLD})}\notag
   \\
   &\leq
   \sigma_{\max}(\Wpi)\norm{\Wp(\yt\!-\!\L\x_{\ML})}\label{eq:WLD_distance_bound_step}
   \\
   &\leq
   \sigma_{\max}(\Wpi)\sigma_{\max}(\Wp)\norm{\yt\!-\!\L\x_{\ML}},\notag
\end{align}
from which~\eqref{eq:WLD_distance_bound} follows. Note that~\eqref{eq:WLD_distance_bound2} and~\eqref{eq:WLD_distance_bound_step} follow because $\norm{\Wp(\yt\!-\!\L\x_{\WLD})}\!\leq\!\norm{\Wp(\yt\!-\!\L\x)}$ for any $\x$.
\end{IEEEproof}

Note that the layer order within the parent set and within the child set is irrelevant. What matters is which layers are selected to form the parent set. $\ILBWLD$ for Gaussian inputs can be used as a criterion for parent layer selection, but the complexity of possible combinations grows as $\Nt\choose \nu$. Alternatively, a less sensitive approach to parent layer selection is to do multiple detection rounds, each time choosing $\nu$ new layers as parents and generating bit LLRs for these parent symbols only.

%
\vspace{-0.05in}
\section{Augmented WLD MIMO Detector}\label{sec:augmented_WLD}
Instead of basing the detection metric in~\eqref{eq:true_metric} on $\H$, we form the augmented vector $\y_{\mathrm{a}}^{\tran}\!\triangleq\![\y^{\tran}~\mbf{0}_{1\times\Nt}]$ and augmented matrix
\begin{align}
    \Ha \triangleq \begin{bmatrix}
        \tfrac{1}{\sqrt{N_0}}\H_{\MbyN}  \\
        \tfrac{1}{\sqrt{\Es}}\I_{\Nt}^{}
      \end{bmatrix}
      \qquad\text{(size $(\Nr\!+\!\Nt)\!\times\!\Nt$)}
      \label{eq:Ha_def}
\end{align}
in a manner analogous to the square-root MMSE~\cite{2000_hassibi_square-root_MMSE}, and reformulate $\mu(\y|\x)$ in~\eqref{eq:true_metric} based on $\Ha$ rather than $\H$ as
\small
\begin{align}
    -\mu(\y|\x)
    &\!=\!
    \tfrac{1}{N_0}\!\norm{\y}^2
    \!-\!\tfrac{2}{\sqrt{N_0}}
    \Re{\![\yh~\mbf{0}]\!\!
        \begin{bmatrix}
            \tfrac{1}{\sqrt{N_0}}\H  \\
            \tfrac{1}{\sqrt{\Es}}\I_{\Nt}^{}
        \end{bmatrix}
        \!\x\!
       }
    \nonumber
    \\
    &\qquad + \xh(\tfrac{1}{N_0}\Hh\H + \tfrac{1}{\Es})\x
    \!-\!\tfrac{1}{\Es}\xh\x
    \nonumber
    \\
    &\!=\!
    \tfrac{1}{N_0}\!\norm{\ya}^2
    \!-\!\tfrac{2}{\sqrt{N_0}}
    \Re{\yah
        \Ha
        \x
        }
    \!+\!\xh\Hah\Ha\x
    \!-\! \tfrac{1}{\Es}\xh\x
    \nonumber
    \\
    &=
    \norm{\tfrac{1}{\sqrt{N_0}}\ya - \Ha\x}^2
    -
    \tfrac{1}{\Es}\xh\x.
    \label{eq:true_distance}\\[-2em]\notag
\end{align}\normalsize
We next expand the squared-distance in~\eqref{eq:true_distance} in terms of the projection matrix $\mbf{P}_{\Ha}\!\triangleq\!\Ha(\Hah\Ha)^{-1}\Hah$ onto the column space of $\Ha$ and its orthogonal complement $\mbf{P}_{\Ha}^{\perp}\!\triangleq\!\I_{\Nr}\!-\!\Ha(\Hah\Ha)^{-1}\Hah$ as
\begin{align}
    \hspace{-0.05in}
    \norm{\!\tfrac{\ya}{\sqrt{N_0}} \!-\! \Ha\x}^2
    &\!\!\!=\!\!
    \norm{\mbf{P}_{\Ha}\!\!(\tfrac{\ya}{\sqrt{N_0}} \!-\! \Ha\x)}^2
    \!\!+\!
    \norm{\mbf{P}_{\Ha}^{\perp}\!\tfrac{\ya}{\sqrt{N_0}}\ya}^2\!\!. \label{eq:true_distance_orth_expansion}
\end{align}
Let $\Qa{}\La{}$ be the thin QL decomposition of $\Ha$:\vspace{-0.05in}
\begin{align}
    \Ha &=   \begin{bmatrix}
                \tfrac{1}{\sqrt{N_0}}\H  \\
                \tfrac{1}{\sqrt{\Es}}\I_{\Nt}^{}
            \end{bmatrix}
        =   \Qa{}\La{}
        =   \begin{bmatrix}
                \Qa{1}  \\
                \Qa{2}
            \end{bmatrix}
            \La{}
        =   \begin{bmatrix}
                \Qa{1}\La{}  \\
                \Qa{2}\La{}
            \end{bmatrix},\label{eq:Ha_QL_decomp_blocks}\\[-2em]\notag
\end{align}
where $\Qa{}$ is an $(\Nr\!+\!\Nt)\!\times\!\Nt$ matrix with orthonormal columns (i.e., $\Qah{}\Qa{}\!=\!\I_{\Nt}$ but not unitary since $\Qa{}\Qah{}\!\neq\!\I_{\Nr+\Nt}$), $\La{}$ is $\NbyN$ lower triangular, and $\Qa{1},\Qa{2}$ are respectively the upper $\MbyN$ and lower $\NbyN$ block matrices of $\Qa{}$. Note that neither the rows nor the columns of $\Qa{1}$ and $\Qa{2}$ are orthonormal. Also, from~\eqref{eq:Ha_QL_decomp_blocks}, it follows that\vspace{-0.05in}
\begin{align}
    \H
    &=
    \sqrt{N_0}\Qa{1}\La{},
    \label{eq:Ha_QL_decomp_Ha_Qa1La}
    \\
    \tfrac{1}{\sqrt{\Es}}\I_{\Nt}^{}
    &=
    \Qa{2}\La{} = \La{}\Qa{2}.
    \label{eq:Ha_QL_decomp_Ha_Qa2La}\\[-2em]\notag
\end{align}
However,~\eqref{eq:Ha_QL_decomp_Ha_Qa1La} is \emph{not} the QL-decomposition of $\H$.~\eqref{eq:Ha_QL_decomp_Ha_Qa2La} implies that $\Qa{2}$ is a lower-triangular matrix proportional to the inverse of $\La{}$, i.e, $\Lai{} \!=\! \sqrt{\Es}\Qa{2}$. Then, from~\eqref{eq:Ha_QL_decomp_blocks} we have\vspace{-0.05in}
\begin{align*}
    \tfrac{1}{N_0}\Hh\H \!+\! \tfrac{1}{\Es}\I_{\Nt}
    &= \Hah\Ha
    = \Lah{}\La{},\\[-2em]\notag
\end{align*}
from which it follows that\vspace{-0.05in}
\small
\begin{align}
    \hspace{-0.06in}
    \norm{\!\tfrac{\ya}{\sqrt{N_0}} \!-\! \Ha\x}^2
    &\!\!\!=\!
    \norm{\La{}\!(\Wmmse\y \!-\! \x)\!}^2
    \!\!\!+\!
    \tfrac{1}{N_0}\!\norm{(\I\!-\!\Qa{}\Qah{})\ya}^2,    \label{eq:true_distance2}\\[-2em]\notag
\end{align}\normalsize
where $\Wmmse$ is the standard MMSE $\NbyM$ filter matrix,
\begin{align}
    \Wmmse
    &\!=\! \Hh  \supsc{[ \H \Hh \!+\! \alpha\I_{\Nr} ]}{-1}
     \!=\!      \supsc{[ \Hh \H \!+\! \alpha\I_{\Nt} ]}{-1} \Hh
    \label{eq:Wmmse_formula_left_right}
    \\
    &\!=\!
    \tfrac{1}{N_0}\!\supsc{(\Hah\Ha\!)}{\,-1}\!\Hh
    \!=\! \tfrac{1}{N_0}\supsc{(\Lah{}\La{})}{\,-1}\!\Hh
    \!=\! \sqrt{\beta}\Qa{2} \Qah{1},
     \label{eq:Wmmse_Ha_La}
\end{align}
with $\alpha \!\triangleq\! \tfrac{1}{\beta} \!=\! \tfrac{N_0}{E_s}$. Substituting~\eqref{eq:true_distance2} back in~\eqref{eq:true_distance}, we obtain
\begin{align}
    \hspace{-0.15in}
   \mu(\y|\x)
   \!=\!
   \tfrac{1}{\Es}\xh\x
   \!-\!
   ||\La{}\!(\Wmmse\y \!-\! \x)||^2
   \!\!\!-\!
   \tfrac{1}{N_0}\!\!\norm{(\I\!-\!\Qa{}\!\Qah{})\ya}^2\!\!.
   \label{eq:true_distance3}
\end{align}

Note that in~\eqref{eq:true_distance3}, the term $\xh\x$ appears explicitly, while tree processing is solely based on $\La{}$ in $||\La{}(\Wmmse\y \!-\! \x)||^2$. We therefore puncture $\La{}$ using an appropriate puncturing matrix $\Wap$ similar to puncturing $\L$ in~\eqref{eq:LP_def} or~\eqref{eq:LP_nu_def1} using $\Wp$. For a given puncturing order $\nu$, we conformally partition $\La{}$ similar to~\eqref{eq:LP_nu_def1} and obtain the partition blocks $\P_{\mathrm{a}}$ of size $\nu\!\times\!\nu$, $\mbf{R}_{\mathrm{a}}$ of size $(\Nt\!-\!\nu)\!\times\!\nu$, and $\S_{\mathrm{a}}$ of size $(\Nt\!-\!\nu)\!\times\!(\Nt\!-\!\nu)$. The resulting punctured augmented matrix $\Lap$ is given by\vspace{-0.05in}\small
\begin{align}
    \Lap
    &\triangleq
    \Wap\La{}
    \label{eq:Lap_def}
    \\
    \Wap
    &\triangleq
    \Dap \diag{\La{}}
        \begin{bmatrix}
            \I_{\nu} & \mbf{0}  \\
            \mbf{0} & \Sai
        \end{bmatrix}
    \!=\!
        \begin{bmatrix}
            \I_{\nu} & \mbf{0}  \\
            \mbf{0} & \mbf{\Sigma}_{\mathrm{a}}\Sai
        \end{bmatrix}
    \label{eq:Wap_formula}
    \\
    \Dap
    &\!=\!
    \supsc{\diag{\La{}}}{-1}\!
    \begin{bmatrix}
        \I_{\nu} & \mbf{0}   \\
        \mbf{0} & \mbf{\Sigma}_{\mathrm{a}}
    \end{bmatrix}
    \label{eq:Dap_formula}
    \\
    \mbf{\Sigma}_{\mathrm{a}}
    &=
    \supsc{\diag{\Sai\!\Saih}}{-\tfrac{1}{2}},\label{eq:Sigma_a}\\[-2em]\notag
\end{align}\normalsize
where $\Dap$ in~\eqref{eq:Dap_formula} is chosen to have $\diag{\Wap\Waph}\!=\!\I_{\Nt}$.

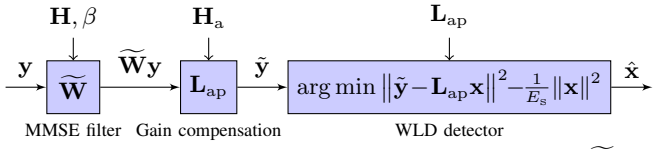
\begin{figure}
  \centering
  \tikzstyle{int}=[draw, fill=blue!20, minimum size=2em]
\tikzstyle{init} = [pin edge={to-,thin,black}]

\begin{tikzpicture}[node distance=2.75cm, auto, >=latex', pin distance = 3mm]]
    \node [int, pin={[init]above:{\small$\H,\beta$}}, label=below:{\scriptsize MMSE
    filter}] (a) {{\small$\Wmmse$\normalsize}};
    
    \node (b) [left of=a,node distance=0.9cm, coordinate] {};
    
    \node [int, pin={[init]above:{\small$\Ha$}}] (c) [right of=a,  label=below:{\scriptsize Gain compensation}, node distance=1.8cm] {{\small$\Lap$\normalsize}};
    
    \node [int, pin={[init]above:{\small$\Lap$}}, label=below:{\scriptsize WLD detector}] (d) [right of=c, node distance=3.2cm] {{\small$\arg\min\norm{\yt\!-\!\Lap\x}^2\!\!-\!\!\tfrac{1}{\Es}\!\norm{\x}^2$\normalsize}};
    
    \node [coordinate] (end) [right of=d, node distance=2.7cm]{};

    \path[->] (b) edge node {{\small$\y$}} (a);
    \path[->] (a) edge node {{\small$\Wmmse\y$\normalsize}} (c);
    \draw[->] (c) edge node {{\small$\yt$\normalsize}} (d);
    \draw[->] (d) edge node {{\small$\hat{\x}$}} (end);

\end{tikzpicture}
 
  \vspace{-0.15in}
  \caption{Block diagram of the AWLD detector, where $\yt\!=\!\Lap\!\Wmmse\!\y$}\label{fig:awld_block_diagram}
\end{figure}

Next, applying $\Wap$ to filter $\La{}(\Wmmse\y \!-\! \x)$ in~\eqref{eq:true_distance3} as\vspace{-0.05in}
\begin{align}
    \norm{\La{}(\Wmmse\y \!-\! \x)}^2
    \xrightarrow{~\Wap~}
    \norm{\Wap\La{}(\Wmmse\y \!-\! \x)}^2, \label{eq:awld_metric_tmp}
\end{align}
and dropping the irrelevant term $\norm{(\I\!-\!\Qa{}\!\Qah{})\ya}^2$, the metric computed by the \emph{augmented} WLD (AWLD) detector corresponding to~\eqref{eq:true_distance3} takes the form\vspace{-0.05in}
\begin{align}
    \muap{\y|\x}
    &\!=\!
    2\Re{\yh\Fap\x} \! - \! \xh\Gap\x \!+\! \tfrac{1}{\Es}\xh\x, \label{eq:awld_metric}
\end{align}
where\vspace{-0.2in}
\begin{align}
    \Fap
    &\triangleq
    \Wmmseh \Gap,~\text{and}  \label{eq:Fap_def}
    \\
    \Gap
    &\triangleq
    \Laph\Lap. \label{eq:Gap_def}
\end{align}
The corresponding AWLD detection model (Fig.~\ref{fig:awld_block_diagram}) becomes\vspace{-0.05in}
\begin{align}
    p_{\mathrm{ap}}(\y|\x)
    &=
    \exp{(2\Re{\yh\Fap\x} \! - \! \xh\Gap\x \!+\! \tfrac{1}{\Es}\xh\x)}. \label{eq:awld_detection_prob}
\end{align}

%
\begin{theorem} Under the same assumptions as Theorem~\ref{thm:ILB_wld}, the AIR of the augmented WLD detector based on~\eqref{eq:awld_detection_prob} with $\Gap,\Fap$ given in~\eqref{eq:Fap_def},~\eqref{eq:Gap_def} respectively, is lower-bounded by\vspace{-0.05in}
\begin{align}
    \ILBAWLD
        \!=\! \Nt\log\Es \!+\! \log \det \!\left( \Laph\Lap \right).
        \label{eq:ILB_AWLD}\\[-2.5em]\notag
\end{align}
\label{thm:ILB_AWLD}
\end{theorem}
\begin{IEEEproof} The lower bound on the AIR of the AWLD detector based on~\eqref{eq:awld_detection_prob} is defined as\vspace{-0.05in}
\begin{align}
    \ILBAWLD
    &= \Ex{\mbf{Y},\mbf{X}}{\log( p_{\mathrm{ap}}(\y|\x) )}
    -\Ex{\mbf{Y}}{\log( p_{\mathrm{ap}}(\y) )}.
    \label{eq:ILB_AWLD_def}
\end{align}
where $p_{\mathrm{ap}}(\y)\!\triangleq\!
\int_{\x\in\C^{\Nt}} p_{\mathrm{ap}}(\y|\x) p(\x)
\,\mathrm{d}\x$ assuming $\x\!\sim\! \mathcal{CN}(\mbf{0},\Es\I_{\Nt})$. The main difference compared to the proof of Theorem~\ref{thm:ILB_wld} is the effect of the term $\tfrac{1}{\Es}\xh\x$ in~\eqref{eq:awld_detection_prob} when evaluating $p_{\mathrm{ap}}(\y)$ under Gaussian densities, which annihilates the effect of the prior density $p(\x)$ to give\vspace{-0.1in}
\begin{align}
    p_{\mathrm{ap}}(\y)
    &\!=\!
    \tfrac{1}{\pi^\Nt \Es^\Nt}\!\!
    \int
        \exp{\left(2\Re{\yh\Fap\x} \!-\! \xh\Gap\x \right)}
        \,\mathrm{d}\x.
        \label{eq:awld_p_ap_tmp}\\[-2em]\notag
\end{align}
After some manipulations, the expectations in~\eqref{eq:ILB_AWLD_def} become
\small
\begin{align*}
    \Ex{\mbf{Y},\mbf{X}}{\log( p_{\mathrm{ap}}(\y|\x) )}
    &=
    \Nt \!-\! \Es\Tr{\Gap} \!+\!2 \Es\Re{\Tr{\Faph\H}}
    \\
    -\Ex{\mbf{Y}}{\log( p_{\mathrm{ap}}(\y) )}
    &=
    \Nt\log\Es \!+\! \log  \det \left( \Gap \right)
    \\
    &\!-\!
    \Tr{ \Faph\left[ \Es\H \Hh + N_0\I  \right] \Fap \Gapi  }
\end{align*}\normalsize
Substituting~\eqref{eq:Gap_def} and~\eqref{eq:Fap_def} for $\Gap$ and $\Fap$, and applying~\eqref{eq:Wmmse_formula_left_right} for $\Wmmse$, then
$\Faph[\Es\H\Hh \!\!+\!\! N_0\I ]\Fap\Gapi \!=\! \Es\Faph\H \!=\! \Es\Gap\!\Wmmse\H$. Also, it is easy to show that
\small
\begin{align}
\Wmmse\H
    &\!=\!
    [\Hh \H \!+\! \a\I_{\Nt}]^{-1} \Hh\H
    =
    \I \!-\! \a[\a\I_{\Nt} \!+\! \Hh\H]^{-1}\label{eq:WmmseH},\\[-1.6em]\notag
\end{align}\normalsize
from which it follows that this matrix product is Hermitian. Therefore, $\Tr{\!\Gap\!\Wmmse\H}\!=\! \Tr{\Gr[\I \!-\! \a\left(\a\I \!+\! \Hh\H\right)^{-1}]}$ is real. Adding the two expectations above results in
\small
\begin{align*}
    \ILBAWLD
    &\!=\!
    \Nt\!\log\Es \!+\! \log  \det\! \left( \Gap \right)
    \!-\! \mathrm{Tr}(\!\Gap[\!\tfrac{1}{\Es}\I \!+\! \tfrac{1}{N_0}\Hh\H]^{-1})
    \!+\! \Nt
    \\
    &\!=\!
    \Nt\!\log\Es \!+\! \log  \det \left( \Gap \right)
    \!-\! \Tr{ \Gap(\Lah{}\La{})^{-1}}
    \!+\! \Nt
    \\
    &\!=\!
    \Nt\!\log\Es \!+\! \log  \det \left( \Gap \right)
    \!-\! \Tr{ \Waph\Wap}
    \!+\! \Nt
\end{align*}\normalsize
from which~\eqref{eq:ILB_AWLD} follows since $\Tr{ \Waph\Wap}\!=\!\Nt$.
\end{IEEEproof}

With the punctured structure of the channel matrix $\Lap$ as given in~\eqref{eq:Lap_def}-\eqref{eq:Dap_formula}, the gap of $\ILBAWLD$ to AWGN capacity can be determined using the following corollary.

%
\begin{corollary} The gap of the AIR of the AWLD detector to AWGN capacity is\vspace{-0.075in}
\begin{align}
    C^{\mathrm{AWGN}} \!-\! \ILBAWLD
    &= \sum_{k=1}^{\Nt-\nu}\!\log
        \left(s_{\mathrm{a}\:kk}^{2}\norm{[\mbf{S}_{\mathrm{a}}^{-1}]_{\bar{k}}}^2+1\right).
        \label{eq:gap_capacity_AWLD}\\[-2em]\notag
\end{align}
where $s_{\mathrm{a}\:kk}^{}$ is the $\nth{k}$ diagonal element of $\mbf{S}_{\mathrm{a}}$ in~\eqref{eq:Wap_formula}, and $[\mbf{S}_{\mathrm{a}}^{-1}]_{\bar{k}}$ is the row vector consisting of the first $k\!-\!1$ elements in row $k$ of $\mbf{S}_{\mathrm{a}}^{-1}$ (excluding the diagonal element). 
\end{corollary}
\begin{IEEEproof} Applying~\eqref{eq:Lap_def}-\eqref{eq:Dap_formula} in~\eqref{eq:ILB_AWLD}, the $\log\det$ term splits and the $C^{\mathrm{AWGN}}\!=\!\log\det( \tfrac{\Es}{N_0}\Hh\H\!+\!\I_{\Nt})$ term emerges.
\end{IEEEproof}

It is worth noting that computing the augmented channel requires simple processing steps comparable to QL decomposition. In particular, matrix inversion is not needed to compute $\Wmmse$ in~\eqref{eq:Wmmse_Ha_La} because the inverse of $\La{}$ is available from~\eqref{eq:Ha_QL_decomp_Ha_Qa2La}. Moreover, following the modular approach of~\cite{2015_mansour_JSP_2x2QAM}, an efficient hardware architecture for an $\NbyN$ AWLD MIMO detector can be constructed from optimized $2\!\times\!2$ MIMO detector cores. Finally, extensions to include soft-input information, imperfect channel estimation effects, and correlated channels are directly applicable based on~\cite{2017_hu_softoutput_AIR}.

%
\section{Modified MIMO Detection Model}\label{s:modified_detection_model}
Instead of working with Euclidean-distance based metrics as in~\eqref{eq:true_metric}, the authors in~\cite{2012_rusek_optimal_channel_short} propose replacing $\H$, $\G$, $N_0$ in~\eqref{eq:true_metric_expanded}
with mismatched parameters $\Hr,\Gr,N_{\mathrm{r}}$ that are subject to AIR optimization. As a result, instead of the true conditional probability in~\eqref{eq:true_detection_prob}, the mismatched model of~\cite{2012_rusek_optimal_channel_short} is\vspace{-0.05in}
\begin{align}
    \mu_{\mathrm{r}}(\y|\x)
    &=
    2\Re{\yh\Fr\x} \! - \! \xh\Gr\x, \label{eq:mismatched_metric_rusek}
    \\
    p_{\mathrm{r}}(\y|\x)
    &=
    \exp{\left( 2\Re{\yh\Fr\x} \! - \! \xh\Gr\x \right)}, \label{eq:mismatched_prob_rusek}
\end{align}
where $N_{\mathrm{r}}$ is absorbed into $\Fr$ and $\Gr$. It is shown in~\cite{2012_rusek_optimal_channel_short} that detectors limited to the Euclidean-based model in~\eqref{eq:true_metric_expanded_sufficient} where $\Gr$ admits a Cholesky factorization proportional to $\Hh\H$ are not optimal from a mutual information perspective because the resulting optimal matrix $\Gr$ to use in~\eqref{eq:mismatched_prob_rusek} may not be positive semi-definite, and hence no such factorization exists. By maximizing the lower bound on the achievable rate based on~\eqref{eq:mismatched_prob_rusek}, the authors in~\cite{2012_rusek_optimal_channel_short} derive an explicit expression for the optimal front-end filter $\Fropt\!=\!(\H\Hh\!+\!\alpha\I)^{-1}\H(\Gr\!+\!\I)$, which is the MMSE filter compensated by the receiver tree processing through $\Gr\!+\!\I$ rather than $\Gr$. Using $\Fropt$, the authors in~\cite{2017_hu_softoutput_AIR} derive an explicit expression for the optimal $\Gr$ so that the tree processing term $\Gropt\!+\!\I$ admits a Cholesky factorization of the form $\mbf{L}^{\opt\dag}\mbf{L}^{\opt}$, such that $\mbf{L}^{\opt}$ has a punctured structured analogous to that of the WLD scheme~\cite{2014_mansour_SPL_WLD}.

In this work, we propose the following modified model\vspace{-0.05in}
\begin{align}
    \mum{\y|\x}
    &=
    2\Re{\yh\F\x} \! - \! \xh\G\x \!+\! \tfrac{1}{\Es}\xh\x, \label{eq:new_metric}\\[-2em]\notag
\end{align}
and $p_{\mathrm{m}}(\y|\x)\!=\!\exp(\mum{\y|\x})$, where tree processing is split into an explicit term $\tfrac{1}{\Es}\xh\x$ separate from $\xh\G\x$ for which $\G$ is subject to optimization, and show that this resulting optimal $\G$ admits a Cholesky factorization. Under such formulation, we show that the optimal front-end filter $\Fopt$ and gain $\Gopt$
that maximize the lower bound on the AIR coincide exactly with those of the augmented WLD detector in~\eqref{eq:Fap_def} and~\eqref{eq:Gap_def}.

%
\begin{theorem} Under the same assumptions as Theorem~\ref{thm:ILB_wld}, the
optimal $\F$ and $\G$ that maximize $\ILB\!=\!\Ex{\mbf{Y},\mbf{X}}{\log( p_{\mathrm{m}}(\y|\x) )} -\Ex{\mbf{Y}}{\log( p_{\mathrm{m}}(\y) )}$ such that $\G$ is positive semidefinite with factor matrices having a punctured structure of order $\nu$ are
\begin{align}
    \Fopt
    &=
    \Wmmseh\Gopt
    \quad\text{and}\quad
    \Gopt
    =
    \Jopth\Jopt, \label{eq:Fopt_Gopt_modified_model}
\end{align}
where $\Wmmse$ is the standard $\NbyM$ MMSE filter matrix in~\eqref{eq:Wmmse_formula_left_right}, and $\Jopt$ is the punctured augmented WLD matrix $\Lap$ given in~\eqref{eq:Lap_def}. Accordingly, the lower bound attained by the AWLD detector in~\eqref{eq:ILB_AWLD} is optimal.
\label{thm:optimal_ILB_modified_model}
\end{theorem}
\begin{IEEEproof} The expectations in the $\ILB$ expression with $p_{\mathrm{m}}(\y|\x)\!=\!\exp(\mum{\y|\x})$ and $p_{\mathrm{m}}(\y)\!=\!\int p_{\mathrm{m}}(\y|\x) p(\x)\,\mathrm{d}\x$ are
\begin{align*}
    \Ex{\mbf{Y},\mbf{X}}{\log( p_{\mathrm{m}}(\y|\x) )}
    &=
    \Nt \!-\! \Es\Tr{\G} \!+\!2 \Es\Re{\Tr{\Fh\H}}
    \\
    -\Ex{\mbf{Y}}{\log( p_{\mathrm{m}}(\y) )}
    &=
    \Nt\log\Es \!+\! \log  \det \left( \G \right)
    \\
    &\!-\!
    \Tr{ \Fh\left[ \Es\H \Hh + N_0\I  \right] \F \Gi  }.
\end{align*}
To determine $\F$ that maximizes $\ILB$, we set the derivative of the terms in the sum of the two expectations involving $\F$ to 0
\begin{align*}
    \frac{\partial}{\partial \F}
    \left(2 \Es\Re{\Tr{\Fh\H}}\!-\!\Tr{ \Fh\![\Es\H \Hh \!+\! N_0\I] \F \Gi}\right)
    \!=\! 0
\end{align*}
from which it follows, after some tedious steps, that\vspace{-0.05in}
\begin{align*}
    \Fopt
    \!=\!
    [ \H \Hh + \tfrac{N_0}{\Es}\I ]^{-1} \H \G
    = \Wmmseh \G.
\end{align*}
Substituting $\Fopt$ back in $\ILB$, and noting that $\Fopth\H\!=\!\G(\Wmmse\H)$ is the product of two Hermitian matrices and hence has real trace, we obtain after further simplifications\vspace{-0.05in}
\begin{align*}
    \ILB
    &\!=\!
    \Nt\log\Es \!+\! \log\det (\G)
    \!-\! \Es \Tr{(\I \!-\! \Wmmse\H)\G} \!+\! \Nt.\\[-2em]
\end{align*}
Using~\eqref{eq:WmmseH}, $\Es(\I \!-\! \Wmmse\H)\!=\!\Es\a[\a\I_{\Nt} \!+\! \Hh\H]^{-1}\!=\!(\Hah\Ha)^{-1}$, where $\Ha$ is defined in~\eqref{eq:Ha_def}. Then\vspace{-0.05in}
\begin{align}\hspace{-0.05in}
    \ILB
    &\!=\!
    \Nt\log\Es \!+\! \log\det (\Jh\J)
    \!-\!  \Tr{(\Lah{}\La{})^{-1}\!\Jh\J} \!+\! \Nt, \label{eq:ILBopt_tilde_J}\\[-2em]\notag
\end{align}
where $\Ha\!=\Qa{}\La{}$ be the QL decomposition of $\Ha$, and $\G\!=\!\Jh\J$ such that $\J$ is a punctured lower triangular matrix of order $\nu$. We next determine $\J$ by maximizing $\ILB$. Assume $\J$ and $\La{}$ are conformally partitioned as\vspace{-0.05in}
\begin{align}
    \J
    &=
    \begin{bmatrix}
        \J_1 &   \\
        \J_2 & \J_3
    \end{bmatrix}
    \qquad\text{and}\qquad
    \La{}
    =
    \begin{bmatrix}
        \mbf{P}_{\mathrm{a}} &   \\
        \mbf{R}_{\mathrm{a}} & \mbf{S}_{\mathrm{a}}
    \end{bmatrix},\label{eq:J_La_partitioned}
\end{align}
where $\J_1,\mbf{P}_{\mathrm{a}}$ are $\nu\!\times\!\nu$ lower triangular,  $\J_3$ is $(\Nt\!-\!\nu)\!\times\!(\Nt\!-\!\nu)$ real diagonal, $\mbf{S}_{\mathrm{a}}$ is $(\Nt\!-\!\nu)\!\times\!(\Nt\!-\!\nu)$ lower triangular, and $\J_2,\mbf{R}_{\mathrm{a}}$ are $(\Nt\!-\!\nu)\!\times\!\nu$ matrices. Note that $\J_3$ is constrained to be a diagonal matrix, not just lower-triangular. Then the trace $\Tr{(\Lah{}\La{})^{-1}\Jh\J}\!=\!\Tr{(\J\Lai{})(\J\Lai{})^{\dag}}\!=\!\normm{\J\Lai{}}_{\mathrm{F}}^2$ in~\eqref{eq:ILBopt_tilde_J} can be computed from $\J\Lai{}$ as\small
\begin{align*}
    \J\Lai{}
    &\!=\!
    \begin{bmatrix}
        \J_1 &   \\
        \J_2 & \J_3
    \end{bmatrix}
    \begin{bmatrix}
        \mbf{P}_{\mathrm{a}}^{-1} &   \\
        -\mbf{S}_{\mathrm{a}}^{-1}\mbf{R}_{\mathrm{a}}\mbf{P}_{\mathrm{a}}^{-1} & \mbf{S}_{\mathrm{a}}^{-1}
    \end{bmatrix}
    \\
    \norm{\J\Lai{}}_{\mathrm{F}}^2
    &\!=\!
    \norm{\J_1\mbf{P}_{\mathrm{a}}^{-1}}_{\mathrm{F}}^2
    \!+\!
    \norm{(\J_2 \!-\! \J_3\mbf{S}_{\mathrm{a}}^{-1}\mbf{R}_{\mathrm{a}})\mbf{P}_{\mathrm{a}}^{-1}}_{\mathrm{F}}^2
    \!+\!
    \norm{\J_3\mbf{S}_{\mathrm{a}}^{-1}}_{\mathrm{F}}^2
\end{align*}\normalsize
Since $\log\det (\Jh\J)$ involves the diagonal terms of $\J_1$ and $\J_3$ only, then $\ILB$ can be optimized for $\J_2$ and $(\J_1,\J_3)$ independently. Setting $\tfrac{\partial}{\partial \J_2}\ILB\!=\!0$, we obtain $\J_2^{\opt} \!=\! \J_3 \mbf{S}_{\mathrm{a}}^{-1} \mbf{R}_{\mathrm{a}}$. Substituting back in $\ILB$, we get 
\begin{align*}
    \ILB
    &\!=\!
    \Nt\log\Es
    \!+\! \log\det (\J_1^{\dag}\J_1)
     \!+\! \log\det (\J_3^{\dag}\J_3)
     \!+\! \Nt
    \\
    & - \norm{\J_1\mbf{P}_{\mathrm{a}}^{-1}}_{\mathrm{F}}^2
      - \norm{\J_3\mbf{S}_{\mathrm{a}}^{-1}}_{\mathrm{F}}^2
\end{align*}
Setting $\tfrac{\partial}{\partial \J_3}\ILB\!=\!0$ and noting that $\J_3$ is real and diagonal, we obtain $2\J_3^{-1}\!-\!2\J_3\diag{\mbf{S}_{\mathrm{a}}^{-1}\mbf{S}_{\mathrm{a}}^{-\dag}}\!=\!0$, from which it follows that $\J_3^{\opt}\!=\!\diag{\mbf{S}_{\mathrm{a}}^{-1}\mbf{S}_{\mathrm{a}}^{-\dag}}^{-1/2}$. Finally, using Lemma~\ref{lem:trace_formula}, it follows that $\J_1^{\opt}\!=\!\mbf{P}_{\mathrm{a}}$. The resulting $\J^{\opt}$ with $\J_1^{\opt},\J_2^{\opt},\J_3^{\opt}$ in place coincides exactly with $\Lap$ given in~\eqref{eq:Lap_def}, and the $\ILB$ attained in~\eqref{eq:ILB_AWLD} is optimal. 
\end{IEEEproof}

%
\begin{lemma} \label{lem:trace_formula}
Let $\U$ and $\V$ be two non-singular square matrices in $\mathcal{C}^{N\times N}$. Let $f(\U,\V) \!=\! \log\det(\U\Uh) \!-\! \Tr{(\U\V)(\U\V)^{\dag}}$ be a real-valued function of complex-valued matrices. Then the optimal $\U$ that maximizes $f$ for a given $\V$ is $\Uopt \!\triangleq\! \argmax_{\U} f(\U,\V)
\!=\! \V^{-1}$, and $f(\Uopt,\V) \!=\! -\sum_{k=1}^{N}\log \tilde{v}_{kk}^2 \!-\! N$, where $\tilde{v}_{kk}$ is the $\nth{k}$ diagonal element of the Cholesky factor of $\V\Vh$.
\end{lemma}
\begin{IEEEproof} Omitted for brevity.
\end{IEEEproof}

%
\textit{Discussion}: We conclude that punctured augmented channel matrices processed by the AWLD detector are optimal in maximizing the lower bound on the achievable information rate. Their structure matches exactly that of AIR-PM, but most importantly, they can be computed using simple QL decomposition followed by Gaussian elimination, resulting in a significant reduction in complexity compared to~\cite{2017_hu_softoutput_AIR}.

%
\section{Efficient Matrix Decomposition Algorithms}\label{s:eff_decomp_algorithm}

%
\subsection{Matrix-Inverse-Free Puncturing via Gaussian Elimination}\label{s:WLD_gaussian_elimination}
Directly inverting $\S$ in~\eqref{eq:Wp_nu_formula} can be avoided if we apply Gaussian elimination to null the elements below the main diagonal of $\S\!=\![s_{ij}]$ in~\eqref{eq:L_nu_partitioned}. Let $\E_k^{}\!\in\!\C^{{(\Nt-\nu)}\times {(\Nt-\nu)}}$ be a Gauss transformation $\E_k^{}\!=\! \I_{\Nt-\nu} \!-\! \subpsc{\bm{\tau}}{k}{} \subpsc{\mbf{e}}{k}{\tran}$, where $\subpsc{\mbf{e}}{k}{\tran}$ is the $\nth{k}$ column of $\I_{\Nt-\nu}$, and $\subpsc{\bm{\tau}}{k}{}$ is the Gauss vector~\cite{2013_golub_matrix} defined as
\begin{align*}
    \subpsc{\bm{\tau}}{k}{\tran}
    \!=\!
    [\underbrace{0,\cdots,0}_{k},\subpsc{\tau}{k+1}{},\cdots,\subpsc{\tau}{\Nt-\nu}{}],
    ~~
    \subpsc{\tau}{i}{}\!=\!\tfrac{s_{ik}}{\skk},
    ~~
    \mbox{\small $i\!=\!k\!+\!1\!:\!\Nt\!-\!\nu$ }.
\end{align*}\\[-1.5em]
Then the operation $\E_k^{}\S$ nulls the entries below the $\nth{k}$ diagonal element in $\S$. Applying this operation repeatedly for $k\!=\!1,\cdots,\Nt\!-\!\nu\!-\!1$ would null all entries in $\S$ below the main diagonal. Grouping these row operations into $\E \!=\! \E_{\Nt-\nu-1}^{}\cdots\E_2^{}\E_1^{}$ results in $\E \mskip1mu \S \!=\!\diag{\S}$, or $\Si\!=\!\supsc{\diag{\S}}{-1}\E$. Setting $\subsc{\mbf{\Sigma}}{\!E} \!=\!
\supsc{\diag{ \mbf{E} \, \supsc{\mbf{E}}{\,\dag} } }{-1/2}$
gives the required product $\mbf{\Sigma}\mskip1mu\Si$ in~\eqref{eq:Wp_nu_formula} inverse-free as $\mbf{\Sigma}\mskip1mu\Si
\!=\! \subsc{\mbf{\Sigma}}{\!E}\E$.

The pseudo-code of the optimized WL decomposition algorithm is shown in Algorithm~\ref{algo:generalized_wld}. It first performs QL decomposition on $\H$ (Algorithm~\ref{algo:qldy}), followed by Gaussian elimination. The code is further optimized to eliminate intermediate matrix multiplication operations to compute the products $\Wp\Qh\y$ and $\Wp\L$. The QL procedure first generates $\yt\!=\!\Qh\y$ as a byproduct to computing $\L$ and $\Q$. Next, starting with $\Wp\!=\!\Q$, the Gaussian elimination loop then immediately applies the same elimination operations to null entries in $\L$ on $\yt$ as well as the corresponding columns of $\Q$.

\begin{algorithm}[hbtp]
\small
\caption{Optimized WL decomposition algorithm}\label{algo:generalized_wld}
%
%
%
\begin{algorithmic}[0]
\FunctionRV{WL}{\H,\y,\nu}{[\Lp, \yp, \Wp]}\Comment{$\H\!:\Nr\!\times\!\Nt$}
\State $[\Q, \L, \yt] \gets \mathsf{QLy}(\H,\y)$
\Comment{\textit{QL decomp.; here $\yt\!=\!\Qh\y$}}
\State $\Wp \gets \Q,~\Lp \gets [\yt~ \L]$
\For{$i\!=\!\nu\!+\!2\!:\!\Nt$}\Comment{\textit{Gaussian elimination}}
    \For{$j\!=\!\nu\!+\!1\!:\!i\!-\!1$}
    \Comment{\textit{col index to puncture}}
        \State{$\alpha \!\gets\! \Lp(i,j\!+\!1)/\Lp(j,j\!+\!1)$}
        \State $\Wp(:,i) \!\gets\! \Wp(:,i) \!-\! \alpha^{\dag}\Wp(:,j)$
        \State $\Lp(i,1\!:\!j\!+\!1) \!\gets\! \Lp(i,1\!:\!j\!+\!1) \!-\! \alpha \Lp(j,1\!:\!j\!+\!1)$
    \EndFor\label{wld_end_for_j}
    \State $\Lp(i,1\!:\!i\!+\!1) \!\gets\! \Lp(i,1\!:\!i\!+\!1)/\norm{\Wp(:,i)}$
    \State $\Wp(:,i) \!\gets\! \Wp(:,i) / \norm{\Wp(:,i)}$
\EndFor\label{wld_end_for_i}
\State $\yp \!\gets\! \Lp(:,1)$
\Comment{$\yp\!:\Nt\!\times\!1$}
\State $\Lp \!\gets\! \Lp(:,2\!:\!\Nt\!+\!1)$
\Comment{$\Lp\!:\Nt\!\times\!\Nt$}
\EndFunctionRV
\end{algorithmic}

\end{algorithm}

\begin{algorithm}[hbtp]
\small
\caption{Optimized QL decomposition algorithm}\label{algo:qldy}
\begin{algorithmic}[0]
\FunctionRV{QLy}{\H,\y}{[\Q, \L, \yt]}
\Comment{$\H\!:\Nr\!\times\!\Nt$}
\State $\Q\gets[\y ~ \H]$ \Comment{\textit{augment $\y$ to $\H$}}
\State $\L \gets \mbf{0}_{\Nt\times{(\Nt+1)}}^{}$
\For{$i\!=\!\Nt\!+\!1\!:\!-1\!:\!2$}\Comment{\textit{col index}}
    \State $\L(i\!-\!1,i) \gets \sqrt{\Q(:,i)^{\dag}\Q(:,i)}$\Comment{\textit{diagonal element}}
    \State $\Q(:,i) \!\gets\! \Q(:,i)/\L(i\!-\!1,i)$
    \For{$j\!=\!i\!-\!1\!:\!-1\!:\!1$}\Comment{\textit{row index}}
        \State $\L(i\!-\!1,j) \!\gets\! \Q(:,i)^{\dag}\Q(:,j)$
        \State $\Q(:,j) \!\gets\! \Q(:,j) \!-\! \L(i\!-\!1,j)\Q(:,i)$
    \EndFor\label{QLDyendforj}
\EndFor\label{QLDyendfori}
\State $\Q \!\gets\! \Q(:,2\!:\!\Nt\!+\!1)$\Comment{$\Q\!:\Nr\!\times\!\Nt$}
\State $\yt \!\gets\! \L(:,1)$\Comment{$\yt\!:\Nt\!\times\!1$}
\State $\L \!\gets\! \L(:,2\!:\!\Nt\!+\!1)$\Comment{$\L\!:\Nt\!\times\!\Nt$}
\EndFunctionRV
\end{algorithmic}

\end{algorithm}

%
\subsection{Eliminating Explicit Computation of MMSE Filter Matrix}\label{s:elim_MMSE_filter}
For the AWLD detector, the MMSE filter matrix $\Wmmse$ in~\eqref{eq:Wmmse_formula_left_right} is needed to compute the metrics in~\eqref{eq:awld_metric_tmp} or~\eqref{eq:awld_metric}. This $\Wmmse$ is to be pre-multiplied with $\Wap\La{}\!=\!\Lap$ and applied to $\y$ in~\eqref{eq:awld_metric_tmp}, or pre-multiplied with $\Laph\Lap$ and then applied to $\y$ in~\eqref{eq:awld_metric}. In either case, working with the quantity $\Wap\La{}\Wmmse$ suffices. On the other hand, equation~\eqref{eq:Wmmse_Ha_La} shows that $\Wmmse$ can be obtain from the QL decomposition of $\Ha$ in~\eqref{eq:Ha_def} as $\sqrt{\beta}\Qa{2} \Qah{1}$ without explicitly inverting $\Ha$. But $\La{}\Qa{2}\!=\!\tfrac{1}{\sqrt{\Es}}\I_{\Nt}^{}$ from~\eqref{eq:Ha_QL_decomp_Ha_Qa2La}, so that $\Wap\La{}\Wmmse\y$ actually reduces to $\tfrac{1}{\sqrt{N_0}}\Wap\Qah{1}\y$. The product $\tfrac{1}{\sqrt{N_0}}\Qah{1}\y$ can be obtained indirectly from the QL decomposition procedure when applied to $\Ha$ and $\tfrac{1}{\sqrt{N_0}}\ya\!=\!\tfrac{1}{\sqrt{N_0}}[\y^{\tran}~\mbf{0}_{1\times\Nt}]$, in addition to generating $\La{}$. Finally, applying $\Wap$ to puncture $\La{}$ can be performed through Gaussian elimination as before, with the elimination operations simultaneously applied to $\tfrac{1}{\sqrt{N_0}}\Qah{1}\y$ to generate the product $\tfrac{1}{\sqrt{N_0}}\Wap\Qah{1}\y$. Therefore, the same $\mathsf{WL}$ algorithm listed in Algorithm~\ref{algo:generalized_wld} when applied to $\Ha$ and $\tfrac{1}{\sqrt{N_0}}\ya$ produces the necessary quantities to compute the distance metrics without any matrix inversion.

%
\section{Simulation Results}\label{s:sim}
\begin{figure}
  \centering
  \includegraphics[scale=0.65]{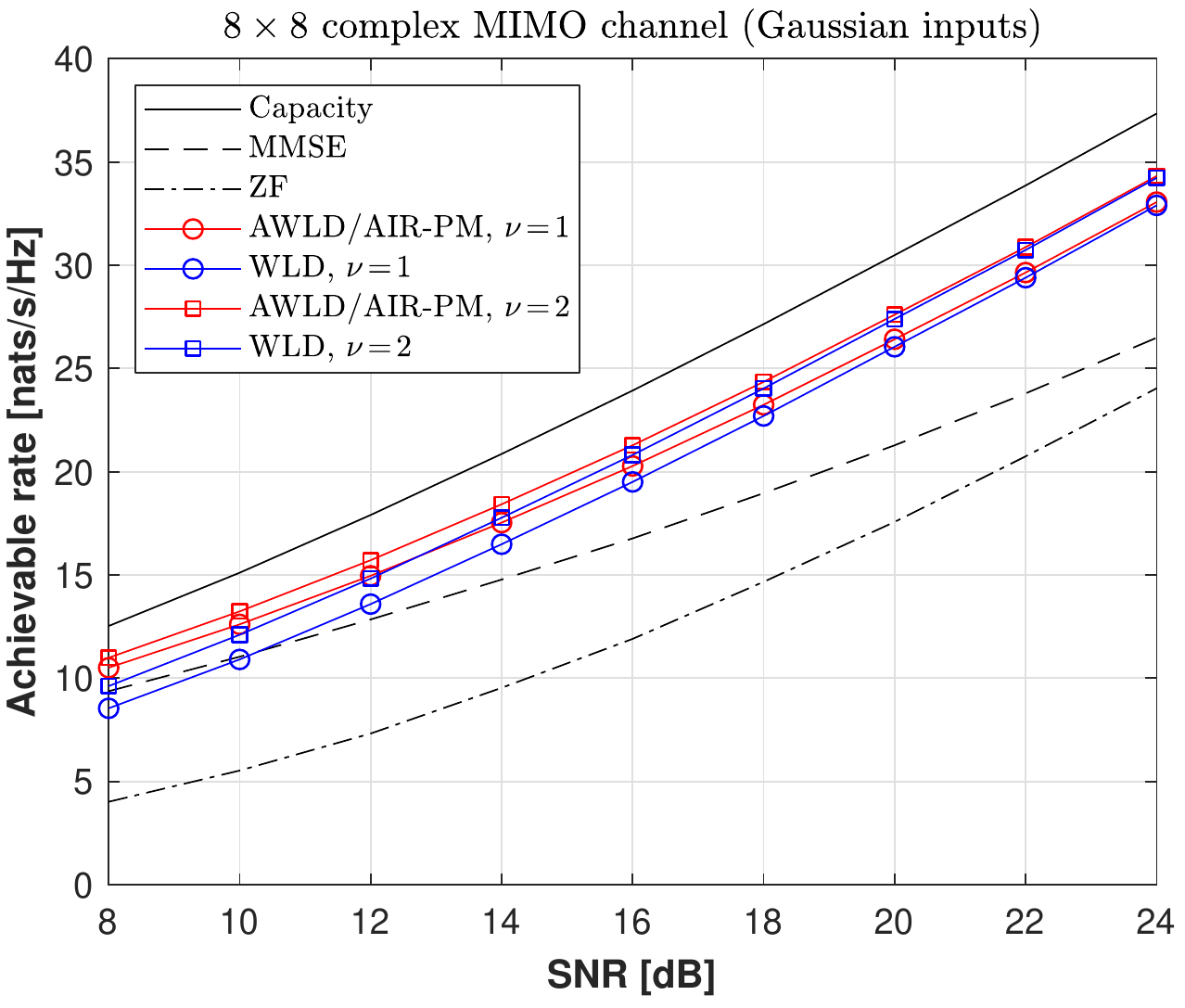}
  \vspace{-0.135in}
  \caption{Comparison of AIRs for $8\!\times\! 8$ MIMO channels with Gaussian inputs.}\label{fig:achievable_rates_8x8}
\end{figure}

\begin{figure}
  \centering
  \includegraphics[scale=0.65]{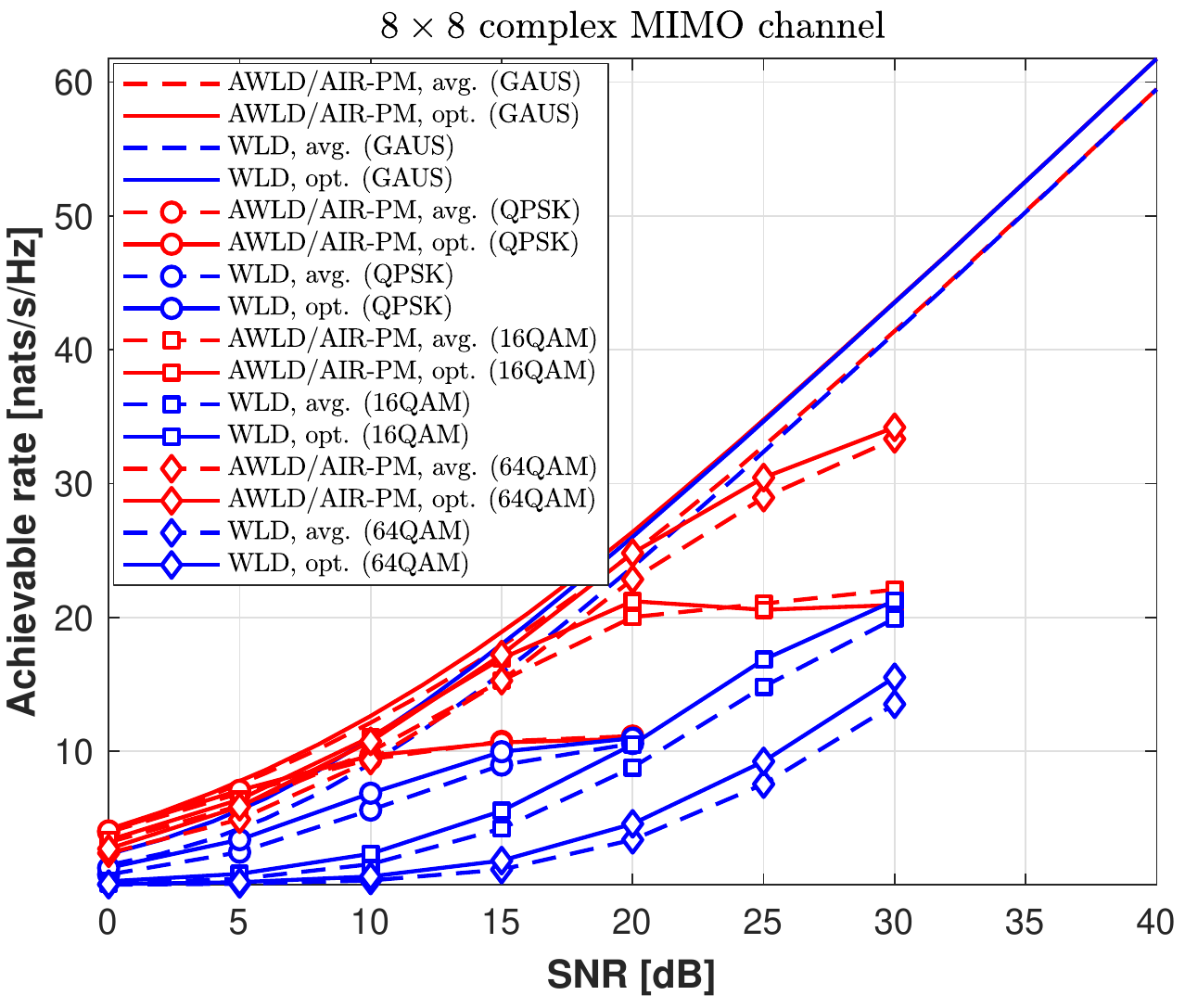}
  \vspace{-0.135in}
  \caption{Comparison of AIRs for $8\!\times\! 8$ MIMO channels with finite inputs.}\label{fig:finite_inputs_rates_8x8}
\end{figure}

In Fig.~\ref{fig:achievable_rates_8x8}, we compare the achievable rates of the proposed AWLD detector against the AIR-PM detector~\cite{2017_hu_softoutput_AIR}, as well as the ZF, MMSE, and WLD~\cite{2014_mansour_SPL_WLD} for $8\!\times\!8$ complex MIMO channels, assuming Gaussian inputs. The AWLD and WLD are simulated for both $\nu\!=\!1$ and $\nu\!=\!2$ configurations. The AWLD attains the same rate as AIR-PM, and for high SNR, the WLD attains a slightly lower rate. On the other hand, Fig.~\ref{fig:finite_inputs_rates_8x8} plots the AIR of AWLD and WLD with $\nu\!=\!1$ for finite QAM constellations. The AWLD achieves higher rates than WLD, especially for 64QAM. Both optimal layer selection and averaging over all layers selected as root are performed.


\begin{figure}[t]
  \centering
  \includegraphics[scale=0.65]{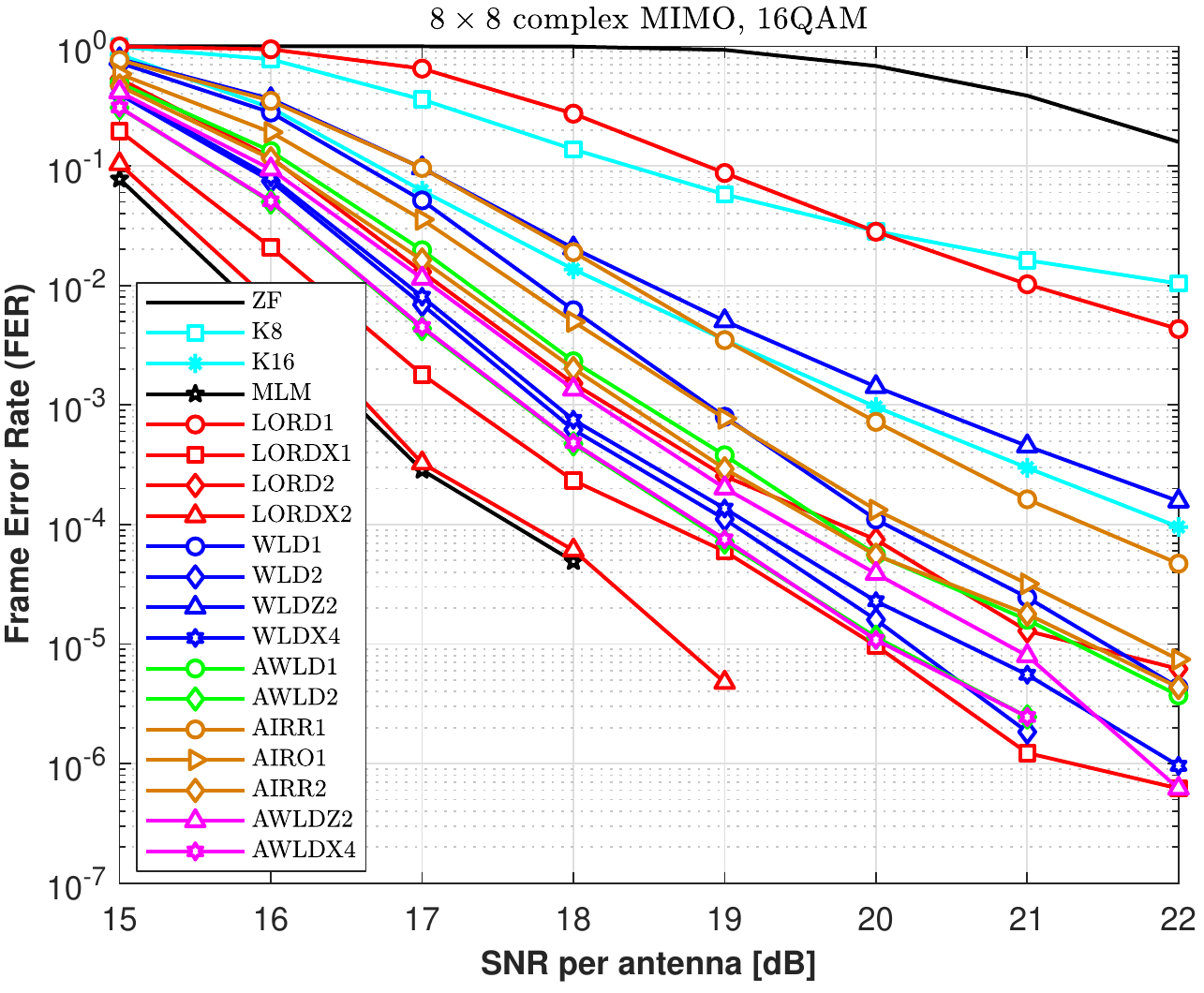}
  \vspace{-0.15in}
  \caption{Frame error-rate of $8\!\times\!8$ complex MIMO channels, 16QAM}\label{fig:fer_plot_8x8_16QAM}
  \vspace{0.05in}
  \includegraphics[scale=0.65]{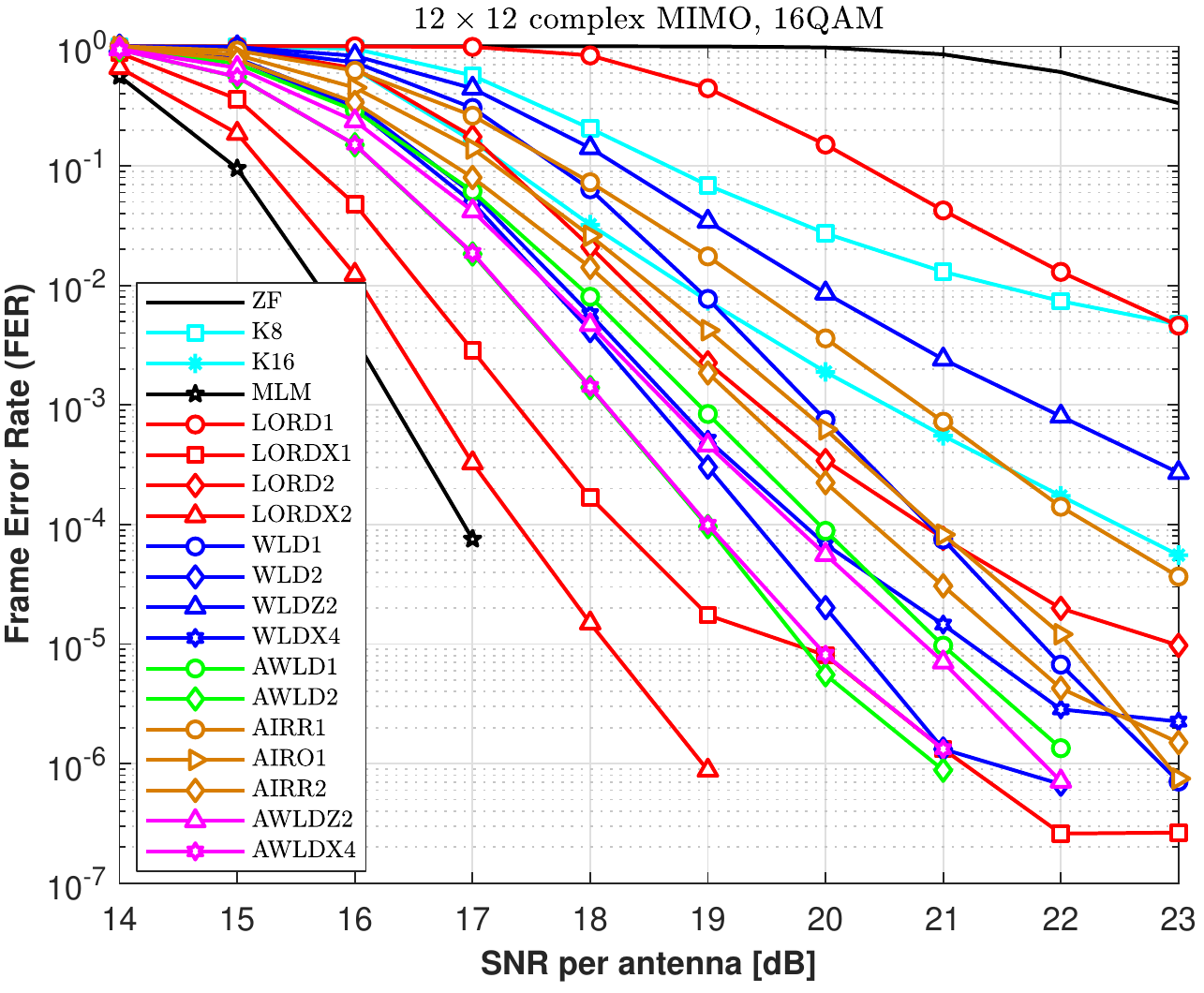}
  \vspace{-0.15in}
  \caption{Frame error-rate of $12\!\times\! 12$ complex MIMO channels, 16QAM}\label{fig:fer_plot_12x12_16QAM}
\end{figure}

In Figs.~\ref{fig:fer_plot_8x8_16QAM}-\ref{fig:fer_plot_12x12_16QAM}, we compare the frame error rate (FER) of the AWLD detector against the Max-Log ML (MLM) sphere decoder with optimized pruning~\cite{2014_sphereP2_mansour}, ZF, K-best~\cite{2006_Wenk_ISCAS}, LORD~\cite{2006_siti_novel_LORD}, WLD~\cite{2014_mansour_SPL_WLD}, and AIR-PM~\cite{2017_hu_softoutput_AIR} detectors for $8\!\times\!8$ and $12\!\times\!12$ MIMO channels with 16QAM, respectively. Max-log approximations for exponential sums are used. An LTE rate-1/2 punctured turbo code of length 1024 is used, and 8 turbo decoder iterations are performed. For LORD, both $\nu\!=\!1,2$ are simulated, and multiple ($\Nt/\nu$) rounds of $\nu$-layer parent selections, QRDs, and ZF-DF steps on the $\Nt\!-\!\nu$ child layers are performed, while tracking the global ML and counter ML hypotheses for all bits. For $\nu\!=\!2$, consecutive layer pairing is done. Similarly for WLD, but without global tracking of global ML and counter ML hypotheses. For AWLD, multiple $\nu$-parent layer selection runs are simulated. For $\nu\!=\!2$, AWLDZ2 does ZF on layer 2, while AWLDX4 searches a window of 4 symbols around the ZF solution on layer 2. AWLDX4 attains better performance compared to the rest, but does not match LORDX1/X2 because it does not benefit from optimizing the metrics globally across the multiple runs as LORD does but at the expense of a significantly higher computational cost.

\begin{figure}
  \centering
  \includegraphics[scale=0.34]{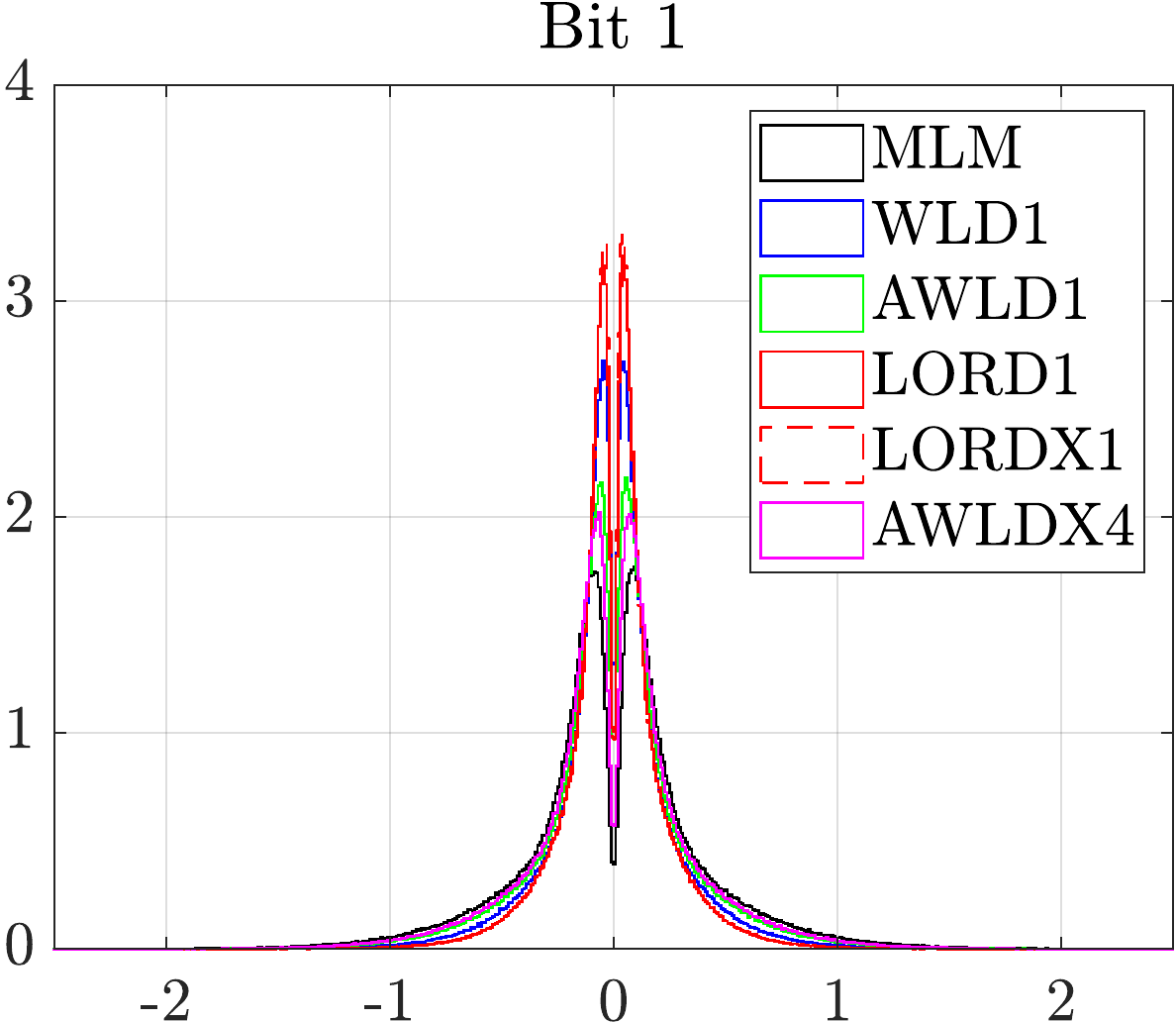}\hspace{-0.0in}
  \includegraphics[scale=0.34]{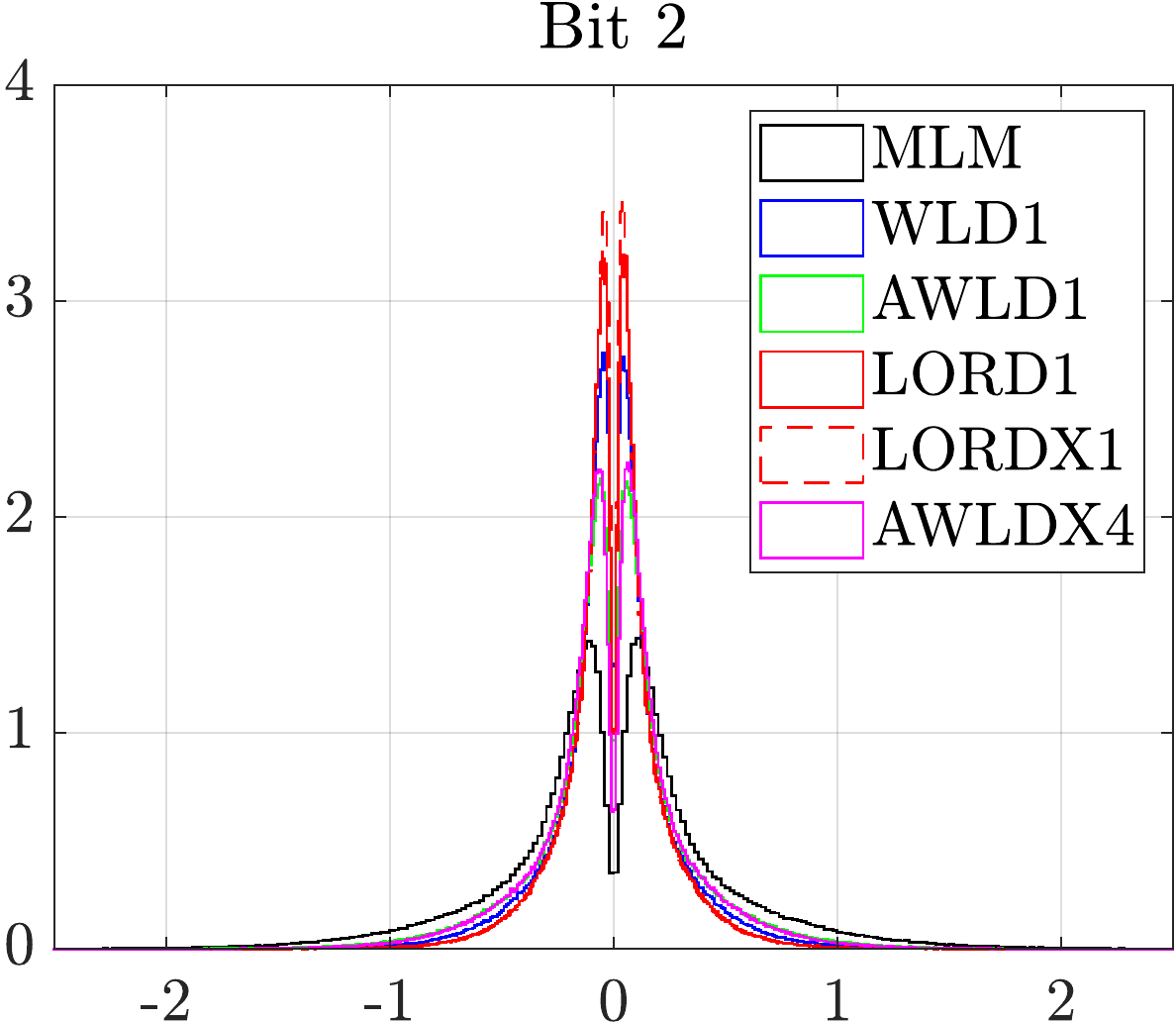}
  \\\vspace{0.025in}
  \includegraphics[scale=0.34]{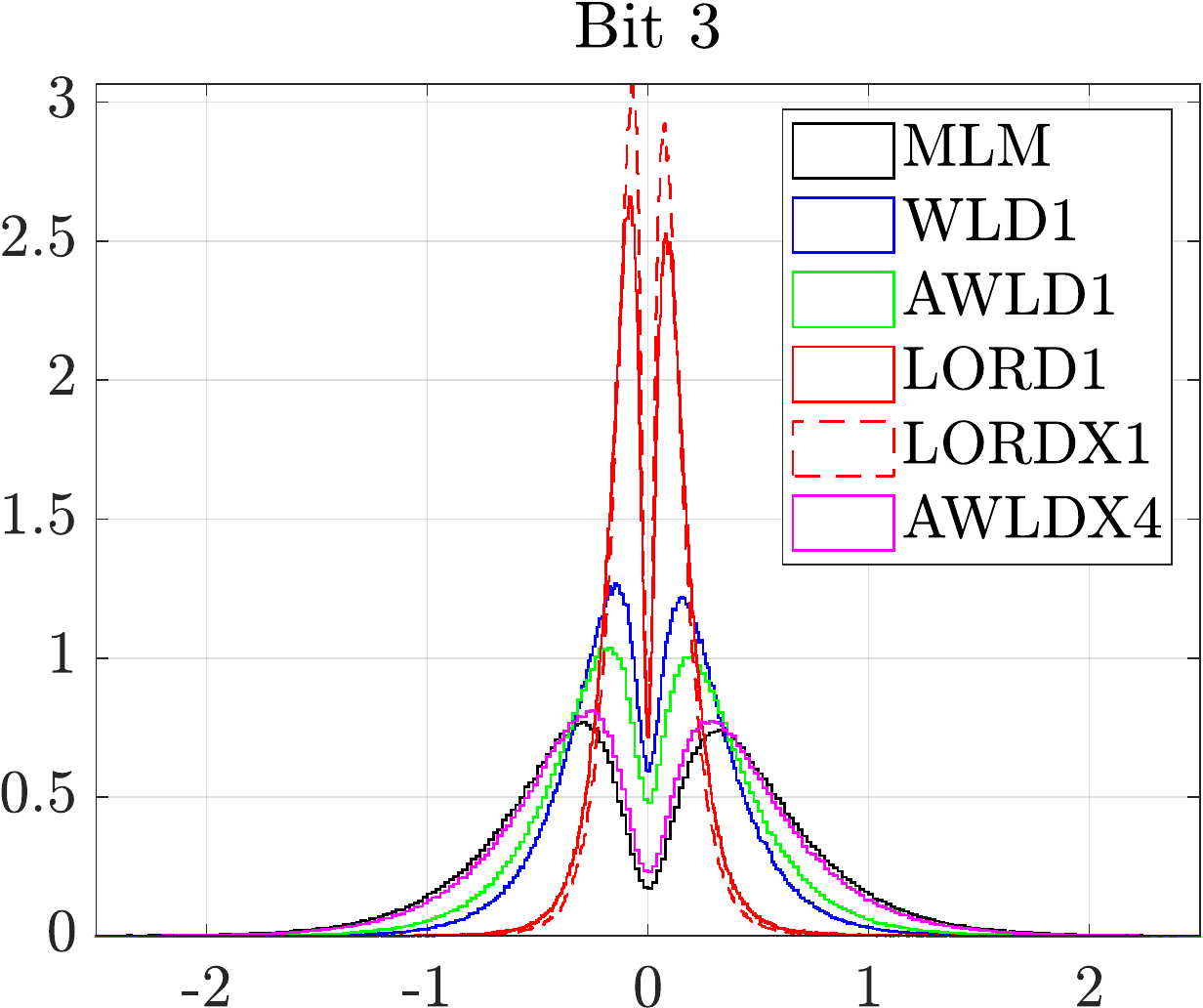}\hspace{-0.0in}
  \includegraphics[scale=0.34]{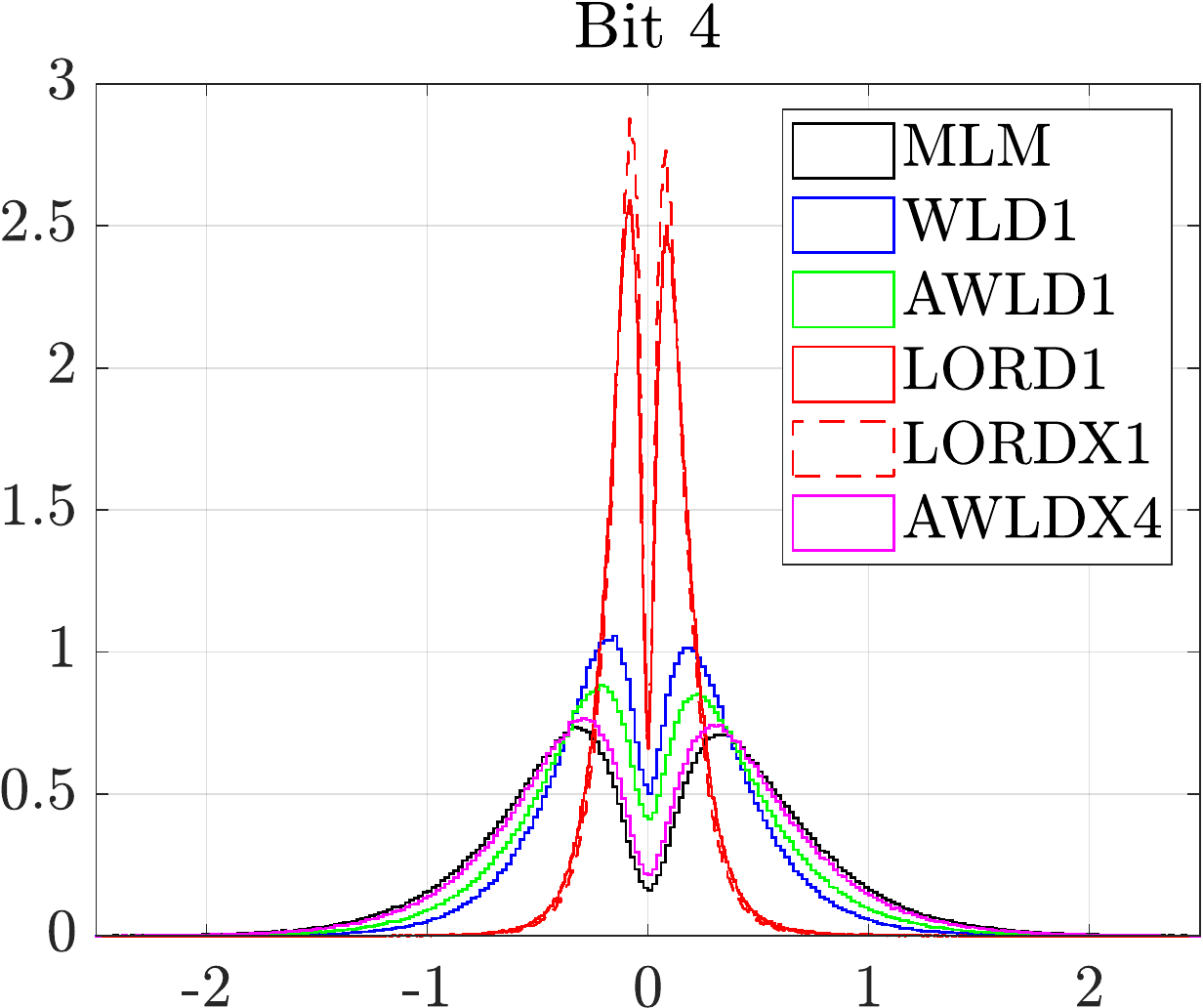}
  \vspace{-0.1in}
  \caption{Distribution of bit LLRs of one symbol: $4\!\times\! 4$ complex MIMO channel, 16QAM, $\text{SNR}\!=\!\unit[20]{dB}$.}
  \label{LLR_bit_1_4_dist_plot_4x4_16QAM}
\end{figure}

Figure~\ref{LLR_bit_1_4_dist_plot_4x4_16QAM} plots the LLR distributions of the first 4 bits in a $4\!\times\! 4$, 16QAM MIMO system at $\text{SNR}\!=\!\unit[20]{dB}$. As shown, AWLDX4 tracks the optimal LLRs very closely.

\bibliographystyle{IEEEtran}
\bibliography{IEEEabrv,bibliography}

\end{document}